\documentclass[12pt]{article}
\usepackage{amsfonts,amssymb,amsmath,fullpage}
\newcommand{\R}{{\mathbb R}}
\newcommand{\Z}{{\mathbb Z}}

\newcommand{\mH}{{\cal Q}}
\newcommand{\mS}{{\mathbb S}}
\newcommand{\T}{{\mathbb T}}
\newcommand{\I}{{\mathbb I}}
\newcommand{\V}{{\mathbb V}}
\newcommand{\mO}{{\mathbb O}}
\newcommand{\D}{{\mathbb D}}
\newcommand{\A}{{\mathbb A}}

\newcommand{\cB}{{\cal B}}

\newcommand{\Fix}{{\rm Fix}}
\newcommand{\ri}{{\rm i}}

\newcommand{\ro}{{\rm o}}
\newcommand{\mbe}{\mbox{\boldmath${\eta}$}}

\newtheorem{definition}{Definition}
\newtheorem{theorem}{Theorem}
\newtheorem{lemma}{Lemma}

\newcommand{\proof}{\noindent{\bf Proof: }}

\newcommand{\qed}{\hfill{\bf QED}\vspace{5mm}}

\begin{document}
\title{Classification and stability\\
of simple homoclinic cycles in $\R^5$}

\author{Olga Podvigina\\
UNS, CNRS, Lab. Lagrange, OCA,\\
BP~4229, 06304 Nice Cedex 4, France;\\
Institute of Earthquake Prediction Theory\\
and Mathematical Geophysics\\
84/32 Profsoyuznaya St, 117997 Moscow, Russian Federation}

\maketitle

\begin{abstract}
The paper presents a complete study of simple homoclinic cycles in $\R^5$.
We find all symmetry groups $\Gamma$ such that a $\Gamma$-equivariant
dynamical system in $\R^5$ can possess a simple homoclinic cycle.
We introduce a classification of simple homoclinic
cycles in $\R^n$ based on the action of the system symmetry group. For systems
in $\R^5$, we list all classes of simple homoclinic cycles. For each class,
we derive necessary and sufficient conditions for asymptotic stability and
fragmentary asymptotic stability in terms of eigenvalues of linearisation
near the steady state involved in the cycle. For any action of the groups
$\Gamma$ which can give rise to a simple homoclinic cycle,
we list classes to which the respective homoclinic cycles belong,
thus determining conditions for asymptotic stability of these cycles.

\end{abstract}

\section{Introduction}\label{sec_intro}

In a generic dynamical system, heteroclinic cycles are of codimension at
least one, but they can be structurally stable in a system with a non-trivial
symmetry group \cite{Kru97}. Hence, a classification of heteroclinic cycles
in a symmetric system can be formulated in terms of symmetry groups and their
actions on $\R^n$, for which heteroclinic cycles can exist.

In this paper we consider
structurally stable homoclinic cycles in a smooth dynamical system
\begin{equation}\label{eq_ode}
\dot{\bf x}=f({\bf x}),\quad f:\R^n\to\R^n,
\end{equation}
equivariant under the action of a non-trivial finite symmetry group $\Gamma$:
\begin{equation}\label{sym_ode}
f(\gamma{\bf x})=\gamma f({\bf x}),\quad\mbox{ for all }
\gamma\in\Gamma\subset{\bf O}(n).
\end{equation}

Let $\xi_1,\ldots,\xi_m\in\R^n$ be hyperbolic equilibria of (\ref{eq_ode})
and $\kappa_j:\xi_j\to\xi_{j+1}$, $j=1,\ldots,m$,
$\xi_{m+1}=\xi_1$, be a set of trajectories from $\xi_j$ to $\xi_{j+1}$.
The union of the equilibria and the connecting trajectories is called
a {\it heteroclinic cycle}. A heteroclinic cycle is {\it structurally stable}
(or robust), if for each $j$ there exists an invariant subspace $P_j$ of
(\ref{eq_ode}) such that $\kappa_j$ belongs to this subspace, and in this
subspace $\xi_{j+1}$ is a sink \cite{am02,Kru97,sot03}. Cycles where all
connections are one-dimensional are called {\it simple heteroclinic cycles}.
If the equilibria are related by a symmetry $\gamma\in\Gamma$,
$\gamma\xi_j=\xi_{j+1}$, then the cycle is called homoclinic.

Heteroclinic cycles often arise in systems related to biology \cite{ml75,hs98},
fluid dynamics \cite{bh80,gh88,pj98} and game theory \cite{ac10}.
Heteroclinic cycles are often responsible for complex and intermittent behaviour
\cite{Kru97}. They may have simple geometric structure but complex local
attraction properties --- heteroclinic cycles
which are not asymptotically stable can attract a positive measure
set in its small neighbourhood \cite{ac10,bra94,dh09,ks94,km04,pos10}.
In \cite{op12} such cycles were called fragmentarily asymptotically stable.

Classifying heteroclinic cycles in $\R^3$ is straightforward.
Heteroclinic cycles in $\R^4$ were categorised into classes A, B and C in
\cite{km95a,km95b,km04}. The symmetry groups giving rise to heteroclinic cycles
of types B and C are known, but this is not the case for cycles of type A.
In \cite{sot03,sot05} a classification of simple homoclinic cycles such that
$\dim P_j=2$ was given for dynamical systems in $\R^4$ and $\R^5$. However,
in these two papers only minimal admissible groups (i.e., the simplest groups
for which homoclinic cycles exist) were found, and only $\Gamma\subset$SO(5)
were considered in $\R^5$.

The theory of asymptotic and fragmentary asymptotic stability is yet
incomplete even for simple heteroclinic cycles. A sufficient condition
for asymptotic stability of heteroclinic cycles was presented in \cite{km95a}.
Necessary and sufficient conditions for asymptotic stability of simple
homoclinic and heteroclinic cycles in $\R^4$ were proven in
\cite{ckms,dh09a,km04}, for fragmentary asymptotic stability of simple
heteroclinic cycles in $\R^4$ in \cite{pa11}. Asymptotic stability of
heteroclinic cycles in some particular systems was studied in
\cite{ac10,bra94,ckms,dh09,dh09a,ks94,km95b,pos10,pd10}.
Necessary and sufficient conditions for asymptotic stability of the
so-called type A cycles were proven in \cite{km95a}, and for asymptotic and
fragmentary asymptotic stability of the so-called type Z cycles in \cite{op12}.

We begin by reminding several examples of homoclinic cycles in $\R^n$ (section
\ref{sec3}) and complete the study of \cite{sot03,sot05} by listing all
possible groups $\Gamma$ acting on $\R^4$, for which simple homoclinic cycles
can arise. Our arguments, based on the quaternion description of the group
SO(4), are much shorter than the ones in \cite{sot03,sot05}. Using
the exhaustive description \cite{mezmezii} of finite subgroups of O(5), we find
in section \ref{sec4} all groups $\Gamma$ acting on $\R^5$ that give rise
to homoclinic cycles in a $\Gamma$-equivariant system. All homoclinic cycles
found in section \ref{sec4} are associated with the cycles listed in section
\ref{sec3}. In section \ref{sec5} we introduce a classification of simple
homoclinic cycles in $\R^n$ based on the isotypic decomposition
of $P_j^{\perp}$ under the action of the symmetry group of $P_j$. Application
of this classification to cycles in $\R^5$ enables us to determine conditions
for asymptotic stability and fragmentary asymptotic stability. Derivation
of conditions for asymptotic stability for two classes of homoclinic cycles
is given in appendices.

\section{Definitions}\label{sec2}

\subsection{Stability}\label{sec_stab}

Let us recall the definitions of various types of asymptotic stability
of an invariant set of system (\ref{eq_ode}). We denote
by $\Phi_t({\bf x})$ the trajectory of (\ref{eq_ode}) that starts
at point $\bf x$. For a set $X$ and a number $\epsilon>0$,
the {\it $\epsilon$-neighbourhood of} $X$ is the set
\begin{equation}\label{ep_nei}
B_{\epsilon}(X)=\{{\bf x}\in R^n:\ d({\bf x},X)<\epsilon\}.
\end{equation}
Let $X$ be a compact invariant set of the system (\ref{eq_ode}). We denote
by $\cB_{\delta}(X)$ its {\it$\delta$-local basin of attraction}:
\begin{equation}\label{del_bas}
\cB_{\delta}(X)=\{{\bf x}\in\R^n:\ d(\Phi_t({\bf x}),X)<\delta\hbox{ for any }
t\ge0\hbox{ and }\lim_{t\to\infty}d(\Phi_t({\bf x}),X)=0\}.
\end{equation}

\begin{definition}\label{def1}
A compact invariant set $X$ is \underline{asymptotically stable}, if for any
$\delta>0$ there exists $\epsilon>0$ such that
$$B_{\epsilon}(X)\subset\cB_{\delta}(X).$$
\end{definition}

\begin{definition}\label{def2}
A compact invariant set $X$ is \underline{fragmentarily asymptotically stable},
if for any $\delta>0$
$$\mu(\cB_{\delta}(X))>0.$$
(Here $\mu$ denotes the Lebesgue measure in $\R^n$.)
\end{definition}

\begin{definition}\label{def3}
A compact invariant set $X$ is \underline{completely unstable}, if there
exists $\delta>0$ such that $\mu(\cB_{\delta}(X))=0$.
\end{definition}

\medskip
We define now various types of stability of a fixed
point $\bf x$ of a map $g:\R^n\to\R^n$ (i.e. $g{\bf x}=\bf x$).

\begin{definition}\label{def11}
A fixed point ${\bf x}\in\R^n$ of a map $g:\R^n\to\R^n$
is \underline{asymptotically stable}, if for any $\delta>0$ there exists
$\epsilon>0$, such that $|{\bf y}-{\bf x}|<\epsilon$ implies
$$|g^k{\bf y}-{\bf x}|<\delta\hbox{ for all }k>0\hbox{ and }
\lim_{k\to\infty}g^k{\bf y}={\bf x}.$$
\end{definition}

\begin{definition}\label{def12}(\cite{op12}, adapted).
A fixed point ${\bf x}\in\R^n$ of a map $g:\R^n\to\R^n$ is
\underline{fragmentarily} \underline{asymptotically stable},
if for any $\delta>0$ the measure of the set
$$V_{\delta}({\bf x})=:\{{\bf y}\in\R^n:\ |g^k{\bf y}-{\bf x}|<\delta
\hbox{ for all }k>0\hbox{ and }\lim_{k\to\infty}g^k{\bf y}={\bf x}\}$$
is positive.
\end{definition}

\begin{definition}\label{def91}
A fixed point ${\bf x}\in\R^n$ of a map $g:\R^n\to\R^n$
is \underline{completely unstable}, if there exists $\delta>0$ such that
$$\mu(V_{\delta}({\bf x}))=0.$$
\end{definition}

\subsection{Heteroclinic cycles}\label{hetc}

Let $\xi_1,\ldots,\xi_m$ be hyperbolic equilibria of the system
(\ref{eq_ode})--(\ref{sym_ode}) with stable and unstable manifolds $W^s(\xi_j)$
and $W^u(\xi_j)$, respectively. Assuming $\xi_{m+1}=\xi_1$, we denote by
$\kappa_j$, $j=1,\ldots,m$, the set of trajectories from $\xi_j$ to $\xi_{j+1}$:
$\kappa_j=W^u(\xi_j)\cap W^s(\xi_{j+1})\ne\emptyset$.

\begin{definition}\label{def4}
A \underline{heteroclinic cycle} is the union of equilibria
$\{\xi_1,\ldots,\xi_m\}$ and their connecting orbits $\{\kappa_1,\ldots,\kappa_m\}$.
\end{definition}

The {\it isotropy group} of a point $x\in\R^n$ is the subgroup of $\Gamma$
satisfying
$$
\Sigma_x=\{\gamma\in\Gamma\ ~:~\ \gamma x=x\}.
$$
The {\it fixed-point subspace} of a subgroup $\Sigma\subset\Gamma$ is the subspace
$$
{\rm Fix}(\Sigma)=\{{\bf x}\in\R^n\ ~: \ \sigma{\bf x}={\bf x}\mbox{ for all }
\sigma\in\Sigma\}.
$$

\begin{definition}\label{def5}
A heteroclinic cycle is \underline{structurally stable (or robust)}, if
for any $j$, $1\le j\le m$, there exist $\Sigma_j\subset\Gamma$ and
a fixed-point subspace $P_j={\rm Fix}(\Sigma_j)$ such that
\begin{itemize}
\item $\xi_{j+1}$ is a sink in $P_j$;
\item $\kappa_j\subset P_j$.
\end{itemize}
\end{definition}

We denote $L_j=P_{j-1}\cap P_j$, by $T_j$ the isotropy subgroup of $L_j$
(evidently, $\xi_j\in L_j$), and by $P_j^{\perp}$ the orthogonal
complement to $P_j$ in $\R^n$.

For a structurally stable heteroclinic cycle, eigenvalues of
$df(\xi_j)$ can be divided into four sets \cite{km95a,km95b,km04}:
\begin{itemize}
\item Eigenvalues with the associated eigenvectors in $L_j$ are called
{\it radial}.
\item Eigenvalues with the associated
eigenvectors in $P_{j-1}\ominus L_j$ are called {\it contracting}.
\item Eigenvalues with the associated eigenvectors in $P_j\ominus L_j$
are called {\it expanding}.
\item Eigenvalues that do not belong to any of the three aforementioned
groups are called {\it transverse}.
\end{itemize}

\begin{definition}\label{def6}(\cite{km04}, adapted).
A robust heteroclinic cycle in $\R^n\setminus\{0\}$
is \underline{simple}, if for any $j$ $\dim(P_{j-1}\ominus L_j)=1$.
\end{definition}

\begin{definition}\label{def7}\cite{op12}
A simple robust heteroclinic cycle is of \underline{type Z}, if for any $j$
\begin{itemize}
\item $\dim P_j=\dim P_{j+1}$;
\item the isotropy subgroup of $P_j$, $\Sigma_j$, decomposes
$P_j^{\perp}$ into one-dimensional isotypic components.
\end{itemize}
\end{definition}

Note that the first condition in the definition \cite{op12} of type Z cycles
is redundant, because it is implied by the simplicity of the cycles:

\begin{lemma}\label{lem00}
Let $X$ be a simple robust heteroclinic cycle, i.e. the respective fixed-point
subspaces satisfy $\dim(P_{j-1}\ominus L_j)=1$ for any $j$, $1\le j\le m$.
Then $\dim P_j=\dim P_{j+1}$ for all $j$, $1\le j\le m-1$.
\end{lemma}

\proof
The dimension of the expanding eigenspace for a steady state $\xi_j$
involved in a heteroclinic cycle can not be less than one, and therefore
$$\dim P_j\ge\dim L_j+1=\dim P_{j-1}.$$
This inequality applied for each $m$ steady states yields
$$\dim P_1\le\dim P_2\le\ldots\le\dim P_m\le\dim P_1$$
(recall that the equilibria are cyclically connected, i.e.,
$\xi_1=\xi_{m+1}$). Here, the leftmost and rightmost values coincide,
and thus all terms in the inequality are equal.
\qed

Since $\dim(P_{j-1}\ominus L_j)=1$ for a simple heteroclinic cycle, all
expanding and contracting eigenspaces are one-dimensional.

\begin{definition}\label{def8}
A simple robust heteroclinic cycle is of \underline{type A$'$},
if for any $j$ the isotypic decomposition of $P_j^{\perp}$ under
the action of $\Sigma_j$ involves only one isotypic component.
\end{definition}

Our A$'$ type cycles are a subset of type A cycles:
\begin{definition} \cite{km95a,km95b,km04}
A simple robust heteroclinic cycle is of \underline{type A}, if for any $j$
\begin{itemize}
\item all eigenvectors, associated with $\lambda^c_j$, $\lambda^t_j$,
$\lambda^e_{j+1}$ and $\lambda^t_{j+1}$, belong to the same isotypic component
in the decomposition of $P_j^{\perp}$ under $\Sigma_j$;
\item all eigenvectors of $df(\xi_j)$, associated with transverse eigenvalues
with positive real parts, belong to the same isotypic component
in the decomposition of $P_j^{\perp}$ under $\Sigma_j$.
\end{itemize}
Here $\lambda^c_j$ and $\lambda^t_j$ denote the contracting and transverse
eigenvalues of $df(\xi_j)$ with the minimum real parts, respectively,
and $\lambda^e_j$ the expanding eigenvalue with the maximum real part.
\end{definition}

Note that if a system depends
on a parameter and the classification involves conditions for eigenvalues, then
on variation of the parameter the type of the cycle can change without any
qualitative change in the overall behaviour of the system.\footnote{For
instance, cycles of class 3-2-[12][3] (see subsection \ref{stab5} and table \ref{tab4})
are not of type A$'$, but for $t_1<t_2$ they are of type A as defined in
\cite{km95a}. Such a cycle is asymptotically stable for $c<e$ and $t_1,t_2<0$.
When $t_1$ exceeds $t_2$, the cycle just ceases to be of type A,
although no bifurcations take place and the stability of the cycle does not
change.} This is not the case, when the classification is
based on the action of the symmetry group, as we propose here.

\begin{definition}\label{def9}
A heteroclinic cycle is called a \underline{homoclinic cycle}, if there
exists a symmetry $\gamma\in\Gamma$ such that for any $1\le j\le m$
$$\gamma\xi_j=\xi_{j+1}.$$
\end{definition}
The symmetry $\gamma$ is called a {\it twist} \cite{am02,sot03}.

\begin{definition}\label{def44}
A \underline{homoclinic\phantom{j}network} is a connected component of the group orbit
of a homoclinic cycle under the action of $\Gamma$.
\end{definition}

In this paper we study stability of simple homoclinic cycles, and we use
the symbols $\xi$, $\kappa$, $P$ and $L$ without subscripts provided this does
not create ambiguity. The radial eigenvalues of $df(\xi)$ are denoted by
$-{\bf r}=\{-r_l\}$, $1\le l\le n_r$, the contracting one by $-c$,
the expanding one by $e$ and the transverse ones by ${\bf t}=\{t_l\}$,
$1\le l\le n_t$. Here $n_r$ and $n_t$ are the numbers of the radial and
transverse eigenvalues, respectively.

\section{Examples of homoclinic cycles}
\label{sec3}

In the section we remind five known examples of homoclinic cycles in $\R^n$
which will be used in the next section in the investigation of simple
homoclinic cycles in $\R^5$. In the last subsection we find all subgroups
$\Gamma\subset$O(4) such that a $\Gamma$-equivariant system
(\ref{eq_ode})--(\ref{sym_ode}) in $\R^4$ can possess a simple homoclinic cycle
under the assumptions that (i) $\dim P=2$ and (ii) the cycle does not belong
to any proper subspace of $\R^4$. We use the homomorphism
$\mH\times\mH\to$SO(4), where $\mH$ is the multiplicative group of unit
quaternions; these notions are reviewed in subsection \ref{quat}.
We end this section
by a table summarising the results of the first and the third subsections.

\subsection{Three simple examples}
\label{exa}

\medskip\noindent
{\it Example 1.} Suppose a system (\ref{eq_ode}) in $\R^3$ is equivariant
with respect to the group $\D_4$ involving two reflections and rotation:
$$s_1:(x_1,x_2,x_3)\to(x_1,-x_2,x_3),\ s_2:(x_1,x_2,x_3)\to(x_1,x_2,-x_3),\
s_3:(x_1,x_2,x_3)\to(-x_1,x_3,x_2).$$
Suppose the system possesses two equilibria in the $x_1$ axis, $\xi_1=(a,0,0)$
and $\xi_2=(-a,0,0)$, that are connected by a trajectory $\kappa_1:\xi_1\to\xi_2$
lying in the plane $(x_1,x_2,0)$. The symmetry $s_3$ permutes the equilibria
and maps the trajectory $\kappa_1$ to $\kappa_2:\xi_2\to\xi_1$ in $(x_1,0,x_3)$.

\medskip\noindent
{\it Example 2.} Let a system (\ref{eq_ode}) in $\R^n$ be equivariant with
respect to the group $(\Z_2)^n\rtimes\Z_n$ acting on $\R^n$
by inversion and cyclic permutation of coordinates. If the system possesses
an equilibrium $\xi_1=(a,0,\ldots)$, then it also has a set of equilibria
$((\pm a,0,\ldots))$ (here double parentheses $((\cdot))$ denote
all cyclic permutations of the quantities listed in the parentheses). Existence
of a connection $\kappa_2:\xi_1\to\xi_2=(0,a,\ldots)$ implies existence
of a homoclinic cycle connecting $n$ equilibria $((a,0\ldots))$, and also
of cycles connecting $2n$ equilibria $((\pm a,0\ldots))$, where only
some combinations of the signs $\pm$ are present in an individual cycle.

\medskip\noindent
{\it Example 3.} The subgroup of $(\Z_2)^n\rtimes\Z_n$ comprised
of orientation-preserving transformations of $\R^n$ is $(\Z_2)^{n-1}\times\Z_n$.
A system (\ref{eq_ode}) with such a symmetry group can possess a homoclinic
cycle comprised of $2n$ equilibria $((\pm a,0\ldots))$.

\medskip
For a given a twist $\gamma$, the symmetry $\gamma\sigma$ is also a twist
for any $\sigma\in T$. The new cycle linked with the twist $\gamma\sigma$
belongs to the same homoclinic network as the cycle with the twist $\gamma$,
but it can involve a different number of equilibria.

\subsection{Quaternions}
\label{quat}

In this subsection we remind some properties of quaternions \cite{conw,pdv}.
A real quaternion is a set of four real numbers, ${\bf q}=(q_1,q_2,q_3,q_4)$.
Multiplication of quaternions is defined by the relation
\begin{equation}\label{mqua}
\begin{array}{ccc}
{\bf q}{\bf w}&=&(q_1w_1-q_2w_2-q_3w_3-q_4w_4,q_1w_2+q_2w_1+q_3w_4-q_4w_3,\\
&&q_1w_3-q_2w_4+q_3w_1-q_4w_2,q_1w_4+q_2w_3-q_3w_1+q_4w_1).
\end{array}\end{equation}
$\tilde{\bf q}=(q_1,-q_2,-q_3,-q_4)$ is the conjugate of $\bf q$, and
$|{\bf q}|^2={\bf q}\tilde{\bf q}=q_1^2+q_2^2+q_3^2+q_4^2$ is the square
of the norm of $\bf q$. ${\bf q}^{-1}=\tilde{\bf q}$ is the inverse of a unit
quaternion $\bf q$. We denote by $\mH$ the multiplicative group of unit
quaternions; obviously, the unity in it is $(1,0,0,0)$.

The four numbers $(q_1,q_2,q_3,q_4)$ can be regarded as Euclidean coordinates
of a point in $\R^4$. The transformation ${\bf q}\to{\bf lqr}^{-1}$
is a rotation in $\R^4$, i.e. an element of the rotation group SO(4). The direct
product $\mH\times\mH$ of two groups $\mH$ is the set of ordered pairs
$({\bf l},{\bf r})$ of elements of $\mH$ equipped with the multiplication
$({\bf l},{\bf r})({\bf l}',{\bf r}')=({\bf ll}',{\bf rr}')$. The mapping
$\Phi:\mH\times\mH\to$SO(4) that relates the pair $({\bf l},{\bf r})$
with the rotation ${\bf q}\to{\bf lqr}^{-1}$ is a homomorphism onto,
whose kernel consists of two elements, $(1,1)$ and $(-1,-1)$; thus
the homomorphism is two to one.

Therefore, a finite subgroup of SO(4) is a subgroup of a product of two
finite subgroups of $\mH$. The group $\mH$ has six finite subgroups:
\begin{equation}\label{finsg}
\renewcommand{\arraystretch}{1.5}
\begin{array}{ccl}
\Z_n&=&\displaystyle{\oplus_{r=0}^{n-1}}(\cos2r\pi/n,0,0,\sin2r\pi/n)\\
\D_n&=&\Z_n\oplus\displaystyle{\oplus_{r=0}^{n-1}}(0,\cos2r\pi/n,\sin2r\pi/n,0)\\
\V&=&((\pm1,0,0,0))\\
\T&=&\V\oplus(\pm{1\over2},\pm{1\over2},\pm{1\over2},\pm{1\over2})\\
\mO&=&\T\oplus\sqrt{1\over2}(\pm1,\pm1,0,0))\\
\I&=&\T\oplus{1\over2}((\pm\tau,\pm1,\pm\tau^{-1},0)),
\end{array}\end{equation}
where $\tau=(\sqrt{5}+1)/2$.

\subsection{Homoclinic cycles in $\R^4$}
\label{clas}

Following \cite{sot03,sot05}, we choose a basis in $\R^4$ such that
$\xi=(0,a,0,0)$ and invariant planes, containing the trajectories that join
the steady states involved in the cycle, are
$$P_1=\gamma^{-1}P=<{\bf e}_1,{\bf e}_2>,\ P_2=P=<{\bf e}_2,{\bf e}_3>,\
P_3=\gamma P=<\cos t {\bf e}_2+\sin t {\bf e}_3,\ \cos s{\bf e}_1+\sin s{\bf e}_4>.$$
Here $t$ is the angle between the fixed point subspaces of two consecutive
equilibria and $s$ is the angle between two consecutive planes. In this basis
the matrix of the twist $\gamma:P_j\to P_{j+1}$~is
\begin{equation}\label{twist}
A=\left(
\begin{array}{cccc}
0&0&\cos s &-\sin s\\
\alpha\sin t&\cos t &0&0\\
-\alpha\cos t&\sin t &0&0\\
0&0&\sin s &\cos s
\end{array}
\right),
\end{equation}
where $\alpha=1$ for orientation-preserving transformations and $\alpha=-1$ for
orientation-reversing ones.

\begin{lemma}\label{lemn5}
Let $X$ be a simple robust homoclinic cycle in $\R^4$, $\dim P=2$,
$\xi$ is an equilibrium involved in the cycle and $T$ is the isotropy group
of $\xi$. Then $T=\Z_2^2$ or $T=\Z_2^3$.
\end{lemma}

\proof
The condition $\dim P=2$ implies that the radial, contracting and expanding
subspaces, $V^r=L$, $V^c$ and $V^e$, respectively, are one-dimensional.
The plane $\gamma^{-1}P=L\oplus V^c$ accommodates two incoming homoclinic
trajectories, and the plane $P=L\oplus V^e$ two outcoming trajectories.
If an element of the group $T$ is of order 3 or more, then $df(\xi)$ has
an eigenspace of dimension 2 or higher. This eigenspace cannot be radial
because $\dim P=2$, and it cannot be transverse because the sum of all
dimensions does not exceed four. If this eigenspace contains the contracting
subspace, then more than two homoclinic trajectories approach $\xi$.
Since in a homoclinic network the numbers of incoming and outcoming trajectories
are equal, we are then lacking dimension(s) for outcoming trajectories. Hence,
the multidimensional eigenspace does not contain the contracting subspace;
similarly it does not contain the expanding one. Therefore, $T$ has no
elements of order higher than two. The group $T$ has an element that acts as
$I$ on $V^c$ and $-I$ on $V^e$. It has another element that acts as $I$ on
$V^e$ and $-I$ on $V^c$; this implies $T=\Z_2^2$ or $T=\Z_2^3$.
\qed

\begin{lemma}\label{lemn6}
Let $s_1$ and $s_2$ be the reflections in $\R^4$ about the planes
$N_1$ and $N_2$, respectively, and $\Phi^{-1}s_1=({\bf l}_1,{\bf r}_1)$
$\Phi^{-1}s_2=({\bf l}_2,{\bf r}_2)$, where $\Phi$ is the homomorphism
defined in the previous subsection. Denote by $({\bf l}_1{\bf l}_2)_1$
and $({\bf r}_1{\bf r}_2)_1$ the first components of the respective quaternion
products. The planes $N_1$ and $N_2$ intersect if and only if
$({\bf l}_1{\bf l}_2)_1=({\bf r}_1{\bf r}_2)_1$.
\end{lemma}

\proof
The planes $N_1$ and $N_2$ intersect if and only if the superposition
$s_1s_2$ is a rotation about a two-dimensional plane by some angle $\alpha$.
The rotation is conjugate to the rotation
about the plane $(0,0,x_3,x_4)$ whose preimage under $\Phi$ is
$$({\bf l}_{\alpha},{\bf r}_{\alpha})=
((\cos\alpha,\sin\alpha,0,0),(\cos\alpha,-\sin\alpha,0,0)).$$
The symmetry $s_1s_2$ is conjugate to this rotation if and only if
there exist quaternions $\bf q$ and $\bf w$ such that
$$\Phi^{-1}s_1s_2=({\bf ql}_{\alpha}{\bf q}^{-1},{\bf wr}_{\alpha}{\bf w}^{-1}).$$
The quaternion ${\bf ql}_{\alpha}{\bf q}^{-1}$ takes the form
$(\cos\alpha,0,0,0)+\sin\alpha\bf v$, where ${\bf v}=(0,v_2,v_3,v_4)$ is a unit
quaternion. On varying $\bf q$, one encounters any such $\bf v$ \cite{pdv}.
${\bf wr}_{\alpha}{\bf w}^{-1}$ takes the same form. Therefore, $s_1s_2$ is
a rotation by $\alpha$ about a two-dimensional plane if and only if
\hbox{$({\bf l}_1{\bf l}_2)_1=({\bf r}_1{\bf r}_2)_1=\cos\alpha$}.
\qed

\begin{theorem}\label{th1}
Consider a $\Gamma$-equivariant system (\ref{eq_ode})--(\ref{sym_ode}) in $\R^4$
possessing a simple homoclinic cycle $X=\{\xi_1,\xi_2,\ldots,\xi_m\}$
linked with the twist $\gamma:\xi_j\to\xi_{j+1}$. Suppose that $\dim P=2$ and
the cycle does not lie in any hyperplane of $\R^4$. Then $\Gamma$ is one of
the following groups:
\begin{itemize}
\item[(i)] $(\Z_2)^4\rtimes\Z_4$ and $m=4$ or 8;
\item[(ii)] $(\Z_2)^3\rtimes\Z_4$ and $m=8$;
\item[(iii)] $\mO\rtimes\Z_4$ and $m=12$ or 24;
\item[(iv)] $\D_{2k}\rtimes\Z_k\rtimes\Z_2$ and $m=4$ or $2k$.
\end{itemize}\end{theorem}

\proof
Lemma \ref{lemn5} implies that $\Sigma_j$ contains the reflection about the plane
$P_j$: if $T=(\Z_2)^2$ then $\Sigma_j=\Z_2$ is comprised of this symmetry and the
identity, otherwise $\Sigma_j=(\Z_2)^2$ and the reflection is its element.

The twist $\gamma$ is or is not a rotation.
We consider the two possibilities separately.

First, suppose $\gamma\in$SO(4). Denote by $s_j$ the reflection about
the plane $P_j$ and by $({\bf l}_j,{\bf r}_j)$ a preimage of $s_j$ under the
homomorphism $\Phi$. It is easy to show that
$$\Phi^{-1}s_1=({\bf l}_1,{\bf r}_1)=((0,1,0,0),(0,1,0,0)),\
\Phi^{-1}s_2=({\bf l}_2,{\bf r}_2)=((0,0,0,1),(0,0,0,-1)).$$
Denote by $({\bf g},{\bf h})$ a preimage of $\gamma$. The reflections $s_1$ and
$s_2$ are conjugate by $\gamma$, i.e. \hbox{$\gamma s_1\gamma^{-1}=s_2$}, or
$${\bf gl}_1{\bf g}^{-1}={\bf l}_2,\quad{\bf hr}_1{\bf h}^{-1}={\bf r}_2.$$
From these relations and the above expressions for $\Phi^{-1}s_i$ we find
$$({\bf g},{\bf h})=((a,b,-a,b),(c,d,c,-d))\hbox{ or }
({\bf g},{\bf h})=((a,b,a,-b),(c,d,-c,d)),$$
where $2a^2+2b^2=2c^2+2d^2=1$.

The group generated by $s_1$ and $\gamma$ is finite if and only if two groups,
one generated by ${\bf l}_1$ and $\bf g$, and the second one generated
by ${\bf r}_1$ and $\bf h$, are finite. The two groups are finite if and only if
$\bf g$ and $\bf h$ are any of the following quaternions (see (\ref{finsg})):
$${1\over\sqrt{2}}(\pm1,0,\pm1,0),\quad{1\over\sqrt{2}}(0,\pm1,0,\pm1),
\quad{1\over2}(\pm1,\pm1,\pm1,\pm1).$$

Recall that a twist is represented by the matrix (\ref{twist}). If $s=\pi k$
for integer $k$, then the homoclinic cycle belongs to a three-dimensional
subspace. Since changing $t\to-t$ or $s\to\pi-s$ yields twists linked with
different cycles that belong to the same homoclinic network, only $t>0$ and
just one of the two values, $s$ or $\pi-s$, are considered. We also restrict
the angle $t$ to take the smallest possible value (for the given group
$\Gamma$), because otherwise an one-dimensional invariant subspace in $P_2$
separates $\xi$ and $\gamma\xi$, making impossible the connection from
$\xi\to\gamma\xi$. Calculating the matrices of the mappings
${\bf q}\to{\bf gqh}^{-1}$ for $\bf g$ and $\bf h$ found above reveals
existence of two twists $\gamma_i$ satisfying these conditions.
For the first one,
$$({\bf g},{\bf h})={1\over2}((0,1,0,1),(1,0,1,0))$$
and the group generated by $s_1$ and $\gamma_1$ is
$\Gamma_1=\D_8\rtimes\Z_2\cong(\Z_2)^3\rtimes\Z_4$ comprised of 32 elements.
For the second one,
$$({\bf g},{\bf h})={1\over2\sqrt{2}}((0,1,0,-1),(1,1,-1,1))$$
and the group generated by $s_1$ and $\gamma_2$ is
$\Gamma_2=\mO\rtimes\Z_4$ is comprised of 192 elements.

Suppose now either of the two groups $\Gamma_i$ is a proper subgroup of
the group of symmetries, $G$, of a dynamical system. Then the system cannot
possess a simple homoclinic cycle linked with the respective twist $\gamma_i$
for the following reasons. If $G\supset\Gamma_1$, then either
$G=\D_{8k}\rtimes\Z_{2k}$ is generated by $s_1$ and
$((0,\cos\beta,0,\sin\beta),(\cos\beta,0,\sin\beta,0))$ for $\beta=\pi/4k$,
or $G=\mO\rtimes\Z_4$. In the first case, the reflection
$((0,\cos2\beta,0,\sin2\beta),(0,\cos2\beta,0,\sin2\beta))$ is an element of
$G$, and by lemma \ref{lemn6} the plane, invariant with respect
to the reflection, intersects with $P_2$. In the second case, the reflection
about the plane intersecting with $P_2$ along the line $x_2=x_3$ is an element
of $G$. Intersection of two invariant planes implies existence
of an one-dimensional invariant subspace in $P_2$ that prohibits the connection
$\xi\to\gamma_1\xi$. If $G\supset\Gamma_2$, then $G=\mO\rtimes\T$
has the sixth-order element ${1\over4}((1,1,1,1),(1,1,1,-1))$
for which ${\bf e}_2$ is invariant, and by lemma~\ref{lemn5}
$(0,a,0,0)$ is not an equilibrium in any homoclinic cycle.

\medskip
Second, suppose $\gamma\notin$SO(4). The preimage, under $\Phi$,
of the reflection $s_3$ about the plane $P_3$ is
$$\Phi^{-1}s_3=({\bf l}_3,{\bf r}_3)=((0,\cos(t+s),\sin(t+s),0),
(0,\cos(t-s+\pi),\sin(t-s+\pi),0)).$$
The reflections $s_1$ and $s_3$ are conjugate by $\gamma^2$.
Denote by $({\bf g},{\bf h})$ a preimage of the square of the twist;
$\gamma^2$ belongs to SO(4) and satisfies
\begin{equation}\label{conj2}
{\bf gl}_1{\bf g}^{-1}={\bf l}_3,\quad{\bf hr}_1{\bf h}^{-1}={\bf r}_3.
\end{equation}
If $t+s\ne k_1\pi/2$ and $t-s\ne k_2\pi/2$, then the only possibilities are
\begin{eqnarray}
&\begin{array}{lll}
{\bf g}&=&(\cos((t+s)/2),0,0,\sin((t+s)/2))\\
&&\hbox{ or }(0,\cos((t+s)/2),\sin((t+s)/2),0),
\end{array}
\label{exg}\\
&\begin{array}{lll}
{\bf h}&=&(\cos((t-s+\pi)/2),0,0,\sin((t-s+\pi)/2))\\
&&\hbox{ or }(0,\cos((t-s+\pi)/2),\sin((t-s+\pi)/2),0)).
\end{array}
\label{exh}
\end{eqnarray}
For $A$ given by (\ref{twist}) and $\alpha=-1$,
\begin{equation}\label{twist2}
A^2=\left(
\begin{array}{cccc}
\cos s\cos t&\cos s\sin t&-\sin^2 s &-\cos s\sin s\\
\cos t\sin t&\cos^2 t &-\cos s\sin t&\sin s\sin t\\
-\sin^2 t&\cos t\sin t &\cos s\cos t&-\sin s\cos t\\
\cos t\sin s&\sin t\sin s&\sin s\cos s &\cos^2 s
\end{array}
\right),
\end{equation}
It easy to check that no $\bf g$ (\ref{exg}) and $\bf h$ (\ref{exh}) yield
a linear transformation ${\bf q}\to{\bf gqh}^{-1}:\R^4\to\R^4$ whose matrix
has this form.

Now suppose $t-s=k_2\pi$. A preimage of the reflection
about $s_3$ is
$$({\bf l}_3,{\bf r}_3)=((0,\pm\cos(t+s),\pm\sin(t+s),0),(0,1,0,0)).$$
For $k_2=0$, (\ref{conj2}) holds for any element ${\bf h}\in$SO(4) and
for $\bf g$ satisfying (\ref{exg}). Under the condition that the matrix of
$\Phi({\bf g},{\bf h})$ is $A^2$ (\ref{twist2}), there are two possibilities
for the square of the twist:
\begin{equation}\label{tw1}
({\bf g},{\bf h})=((\cos((t+s)/2),0,0,\sin((t+s)/2),(\cos(\pi-(t+s)/2),0,0,\sin(\pi-(t+s)/2)))
\end{equation}
or
\begin{equation}\label{tw2}
({\bf g},{\bf h})=((0,\cos((t+s)/2),\sin((t+s)/2,0),(0,\cos(\pi-(t+s)/2),\sin(\pi-(t+s)/2),0)).
\end{equation}
The group generated by $({\bf l}_1,{\bf r}_1)$ and $({\bf g},{\bf h})$
is finite if $t+s=4\pi/k$.
In fact, the group generated by (\ref{tw1}) and $s_1$, and the group generated
by (\ref{tw2}) and $s_1$, are identical. Two cycle linked with the square
of the twist (\ref{tw1}) belongs to the same homoclinic network as the cycle
linked with (\ref{tw2}). Note that the order of (\ref{tw1}) is $k$ and the
order of (\ref{tw2}) is two. The group generated by (\ref{tw2}) and $s_1$ is
$\D_{2k}\rtimes\Z_k$. The group $\Gamma$ is a product of $\D_{2k}\rtimes\Z_k$
with $\Z_2$, where $\Z_2$ is generated by an orientation-reversing symmetry.

We checked directly that for the pair $t+s\ne k_1\pi/2$ and $t-s=\pi/2+k_2\pi$,
as well as for the pair $t+s=k_1\pi/2$ and $t-s=k_2\pi/2$, no twists of the form
(\ref{twist2}) are possible.

Suppose now a dynamical system has a group of symmetries $G_0\rtimes\Z^2$, where
$\D_{2k}\rtimes\Z_k$ is a proper subgroup of $G_0$.
$\D_{2k}\rtimes\Z_k$ is a subgroup of $\D_{2rk}\rtimes\Z_{rk}$
for any $r\ge2$; for $k>2$ it is not a subgroup of any other finite subgroup of SO(4).
The group $\D_{2rk}\rtimes\Z_{rk}$ contains a reflection with respect to a plane
that intersects with the plane $P_2$, this prohibiting connections
$\xi\to\gamma\xi$. For $k=2$, the two remaining possibilities are
$G_0=\D_8\rtimes\Z_2\cong(\Z_2)^3\rtimes\Z_4$ or $G_0=\mO\rtimes\Z_4$.
In the first case, a system with the symmetry $G_0\rtimes\Z^2\cong(\Z_2)^4\rtimes\Z_4$
can possess a homoclinic cycle linked with the twist $\gamma$ (such a system
is an instance of example 2 in subsection \ref{exa}). The reflection about
the plane intersecting with $P_2$ along the line $x_2=x_3$ belongs to
$\mO\rtimes\Z_4$. Thus the line $x_2=x_3$ is an invariant subspace
of the dynamical system, that prohibits the connection $\xi\to\gamma\xi$
in the second case.
\qed

Table \ref{tab1} summarises the results of this section.

\begin{table}[t]
\begin{center}
$$
\begin{array}{llllll}
n & \Gamma & |\Gamma| & N &\Sigma & \mbox{twist}\\
\hline
3&\D_4& 8 & 2 & \Z_2 & g\in\D_4,\ g^2=e\hbox{ or }g^4=e\\
n & (\Z_2)^n\rtimes\Z_n & n2^n & n\hbox{ or }2n & (\Z_2)^{n-2} & g\hbox{ or }sg;\ g\in\Z_n,\ s\in\Z_2\\
n & (\Z_2)^{n-1}\rtimes\Z_{n} & n2^{n-1} & 2n &(\Z_2)^{n-3} & sg;\ g\in\Z_{n},\ s\in\Z_2\\
4 & \mO\rtimes\Z_4 & 192 & 12\hbox{ or }24 & \Z_2 & g\hbox{ or }sg;\ g\in\mO,\ s\in \Z_4\\
4 & \D_{2k}\rtimes\Z_k\rtimes \Z_2,k\ge2 & 8k^2 & 4\hbox{ or }2k& \Z_2 & sg;g\in\D_{2k},\ s\in\Z_2,\
g^2=e\hbox{ or }g^k=e
\end{array}
$$
\end{center}
\caption{\label{tab1}
Examples of symmetry groups $\Gamma$ of the system (\ref{eq_ode})--(\ref{sym_ode})
in $\R^n$ for which simple homoclinic networks are possible. The fourth
column shows the number of equilibria for the homoclinic cycles
involved in the network, and the last column the respective twists.}
\end{table}

\section{Homoclinic cycles in $\R^5$}
\label{sec4}

In this section we
list all symmetry groups $\Gamma$ such that a $\Gamma$-equivariant system
(\ref{eq_ode})--(\ref{sym_ode}) in $\R^5$ can possess a simple homoclinic cycle.

\medskip
\begin{theorem}\label{thmezi} \cite{mezmezii}.
Let $\Gamma$ be a finite subgroup of the orthogonal group {\rm SO(5)} or {\rm O(5)}.
Then at least one of the following statements is true:
\begin{itemize}
\item[(i)] $\Gamma$ is conjugate to a subgroup of $W=(\Z_2)^4\rtimes\mS_5$
or $\widetilde W=(\Z_2)^5\rtimes\mS_5$;

\item[(ii)] $\Gamma$ is isomorphic to $\A_5$, $\mS_5$, $\A_6$ or $\mS_6$,
or to the product of one of these groups with $\Z_2=<-I>$;

\item[(iii)] $\Gamma$ is conjugate to a subgroup of {\rm O(4)}$\times${\rm O(1)}
or {\rm O(3)}$\times${\rm O(2)}.
\end{itemize}\end{theorem}

Note that these possibilities are not mutually exclusive.
Here the semidirect product \hbox{$\widetilde W=(\Z_2)^5\rtimes\mS_5$ acts on $\R^5$}
by inversion and permutation of coordinates, and $W=(\Z_2)^4\rtimes\mS_5$ is
its subgroup, of index two, consisting of orientation-preserving elements.
The symmetric group $\mS_n$ acts on $\R^n$ by permutation of coordinates,
and also on its subspace isomorphic $\R^{n-1}$ in which the sum of all
coordinates is zero.

\begin{lemma}\label{th9}
Consider a $\Gamma$-equivariant system (\ref{eq_ode})--(\ref{sym_ode})
in the subspace of $\R^6$ in which the sum of all coordinates is zero
(isomorphic to $\R^5$), where $\Gamma$ is one of the following: $\mS_6$,
$\A_6$, $\mS_6\times\Z_2$ or $\A_6\times\Z_2$ for $Z_2=<-I>$. Such a system
cannot possess a simple homoclinic cycle.
\end{lemma}

\proof
We begin by showing that if $\Gamma=\mS_6$ or $\Gamma=\A_6$, then the system can
only possess a homoclinic cycle that is not simple. Suppose the system has
an equilibrium \hbox{$\xi_1=(a,a,a,-a,-a,-a)$} with an unstable eigenspace
$V_1=(b,c,d,0,0,0)$ such that\break$b+c+d=0$ and a stable eigenspace $V_2=(0,0,0,e,f,g)$,
$e+f+g=0$. The symmetry $s:{\bf x}\to(x_4,x_5,x_6,x_1,x_2,x_3)$ maps $\xi_1$ to
$\xi_2=(-a,-a,-a,a,a,a)$. In the hyperplane $(b,c,d,a,a,a)$, where $b+c+d+3a=0$,
the steady state $\xi_1$ is unstable and $\xi_2$ is stable; therefore,
the heteroclinic connection $\kappa_1=W^u(\xi_1)\cap W^s(\xi_2)$ is structurally
stable. The symmetry $s$ maps $V_1$ to $V_2$, where the connection
$s\kappa_1:\xi_2\to\xi_1$ exists. Note that both $\kappa_1$ and $\kappa_2$ are
two-dimensional manifolds, and therefore the cycle is not simple.
This connection does not exist if $-I\in\Gamma$, because the symmetry $-I$
maps $\xi_1$ to $\xi_2$ and the hyperplane $(b,c,d,a,a,a)$ is $-I$-invariant.
Hence the equilibria are simultaneously stable or unstable in this subspace.

\medskip
To show that no other homoclinic cycles exist, we consider a fixed-point
subspace of dimension not exceeding three for a group listed in the condition
of the theorem and prove that no simple homoclinic cycles exist that connect
equilibria of a given symmetry type. First note that two equilibria,
$(a,-a,0,0,0,0)$ and $(0,0,a,-a,0,0)$, might be connected within
the subspace $(b,-b,c,-c,0,0)$, but existence of the invariant subspace
$(b,-b,b,-b,0,0)$ separating the two equilibria prohibits this. Second, steady
states $(a,a,b,b,b,b)$ and $(b,b,a,a,b,b)$, for $a+2b=0$, belong to the plane
$(c,c,d,d,e,e)$, where $c+d+e=0$. The plane is invariant under
the transformation ${\bf x}\to(x_3,x_4,x_1,x_2,x_5,x_6)$ that interchanges
the two steady states; hence within the plane they are simultaneously stable
and unstable and no robust homoclinic connection from one steady state
to another exists. Proofs on non-existence of homoclinic connections between
any other pair of steady states reduce to the two above arguments.
\qed

\begin{definition}\label{defn1}
Consider a system (\ref{eq_ode})--(\ref{sym_ode}) possessing a robust
$\gamma$-symmetric heteroclinic cycle $X=\{\xi_1,\ldots,\xi_m\}$ (i.e.
$\gamma(\xi_j)=\xi_{j+k}$ holds for all $j$, $1\le j\le m$, and some $k$
independent of $j$; for a homoclinic cycle, $k=1$). The \underline{symmetry
subgroup} of $X$, $G(X)$, is the subgroup of $\Gamma$ generated by $\Sigma_j$,
$1\le j\le m$, and $\gamma$.
\end{definition}

\begin{definition}\label{defn2}
Consider a system (\ref{eq_ode})--(\ref{sym_ode}) possessing a robust
heteroclinic cycle $X$,\break$X=\{\xi_1,\ldots,\xi_m\}$.
The \underline{essential subspace} of $X$, $V_{\rm Ess}(X)$, is the smallest
$G(X)$-invariant subspace of $\R^n$ which contains all contracting and expanding
eigenvectors of the equilibria involved in the cycle.
\end{definition}

\begin{definition}\label{defn4}
Consider a $\Gamma$-equivariant system (\ref{eq_ode})--(\ref{sym_ode})
possessing a robust $\gamma$-symmetric heteroclinic cycle
$X=\{\xi_1,\dots,\xi_m\}$. The \underline{essential subgroup} of $X$,
$G_{\rm Ess}(X)$, is the group $G(X)/\Sigma_{V_{\rm Ess}(X)}$.
\end{definition}

\begin{lemma}\label{lemn7}
Suppose a $\Gamma$-equivariant system (\ref{eq_ode})--(\ref{sym_ode}) in $\R^5$
possesses a simple homoclinic cycle $X$ with an equilibrium $\xi$
and $T$ is the isotropy group of $\xi$.
Consider the isotypic decomposition of $\R^5$ under the action of $T$:
\begin{equation}\label{isr}
\R^5=W_1\oplus\ldots\oplus W_K.
\end{equation}
Suppose $W_1$ is the isotypic component in which $T$ acts trivially,
$W_2$ contains the contracting eigenvector and $W_3$ contains
the expanding eigenvector.
Then $\dim W_2=\dim W_3$ and $\dim W_k\le 2$ for any $2\le k\le K$.
\end{lemma}

The proof is similar to that of lemma \ref{lemn5} and is not presented.

\begin{lemma}\label{th8}
Suppose a $\Gamma$-equivariant system (\ref{eq_ode})--(\ref{sym_ode})
in $\R^5$ possesses a simple homoclinic cycle which does not belong to
any proper subspace of $\R^5$. Suppose $\Gamma\subset\widetilde W$,
$\Gamma\not\subset${\rm O(4)}$\times${\rm O(1)} and $\Gamma\not\subset${\rm O(3)}$\times${\rm O(2)}.
Then
\begin{itemize}
\item $\Gamma=(\Z_2)^5\rtimes\Z_5$ or $\Gamma=(\Z_2)^4\rtimes\Z_5$;
\item $\dim P$=2, 3 or 4 and $\Sigma=(\Z_2)^3$, $(\Z_2)^2$ or $\Z_2$.
\end{itemize}\end{lemma}

\proof
Since $\Gamma$ is a subgroup of $\widetilde W=\mS_5\rtimes(\Z_2)^5$, we have
$\Gamma=G\rtimes R$, where $R\subset(\Z_2)^5$ is the subgroup of $\Gamma$
comprised of reflections and $G=\Gamma/R$. For the sake of simplicity we assume
that all symmetries in $G$ are just coordinate permutations. If this is
not the case, the proof is similar and is not presented.

Since the group $\Gamma$ is not a subgroup of O(4)$\times$O(1) or
O(3)$\times$O(2), it has a fifth-order element which is a cyclic permutation
of coordinates. Without any loss of generality we assume that it is the
permutation ${\bf x}\to(x_2,x_3,x_4,x_5,x_1)$.

Suppose $\dim P=2$. We begin by showing that equilibria involved in the cycle
are located on coordinate axes. Let us assume the converse. An equilibrium
belongs to an one-dimensional subspace of fixed points for a subgroup
of $\Gamma$. Such subspaces are
$$(a,0,0,0,0),\ (a,a,0,0,0),\ (a,a,a,0,0),\ (a,a,a,a,0)\ \hbox{or~}(a,a,a,a,a).$$
We need to find $\Gamma=G\rtimes R\subset\mS_5\rtimes(\Z_2)^5$ for which such
a subspace (not of the first type) is a fixed-point subspace of a subgroup
of $\Gamma$. The permutation
${\bf x}\to(x_2,x_3,x_4,x_5,x_1)$ generates $\Z_5=<(1,2,3,4,5)>$. By our
assumption, the permutation belongs to $\Gamma$; hence $G$, a subgroup
of $\mS_5$, contains $\Z_5$. Only five such subgroups exist:
$$\Z_5,\ \D_5=<(1,2,3,4,5),(2,5)(3,4)>,\ \hbox{GA(1,5)}=<(1,2,3,4,5),(1,2,3,4)>,$$
$$\A_5=<(1,2,3,4,5),(1,2,3)>\hbox{ and }\mS_5.$$
The subspace $(a,a,0,0,0)$ is invariant, if $G=\D_5$ or $G=\mS_5$.
If $G=\D_5$, then the system does not have an invariant plane $P$ that contains
$(a,a,0,0,0)$ and its symmetric copy. If $G=\mS_5$, then by lemma \ref{lemn7}
an equilibrium $(a,a,0,0,0)$ is not involved in a homoclinic cycle.
Similarly, lemma \ref{lemn7} prohibits homoclinic cycles involving
$(a,a,a,0,0)$, $(a,a,a,a,0)$ or $(a,a,a,a,a)$.

Thus we have established that equilibria involved in the cycle are located
on coordinate axes. The twist $\gamma$ is a cyclic permutation of coordinates
by the assumption that the cycle does not belong to any proper subspace
of $\R^5$. The group $\Sigma$ acts trivially on the coordinate plane $P$ and
non-trivially on $P^{\perp}\cong\R^3$. If $\Sigma$ contains a symmetry $r$
that acts on $P^{\perp}$ as $-I$, then $R=(\Z_2)^5$, because
$\gamma^kr\gamma^{-k}$ for $1\le k\le 5$ generate all reflections in $\R^5$.
In this case $\Sigma=(\Z_2)^3$.
If $\Sigma$ does not contain $-I$, then $\Sigma=(\Z_2)^2$ is generated
by symmetries changing signs of two coordinates, and $R=(\Z_2)^4$ is comprised
of symmetries changing signs of an even number of coordinates.

Thus it is left to show that $G=\Z_5$. Assume the converse, i.e. that $\Z_5$
is a proper subgroup of $G$. Only four subgroups of $\mS_5$ contain $\Z_5$, they are
$\D_5$, GA(1,5), $\A_5$ and $\mS_5$. By lemma \ref{lemn7}, the group $G$ is not
GA(1,5), $\A_5$ or $\mS_5$, because
in these cases $T$, the isotropy subgroup of $\xi$, contains an element
permuting three or four coordinates. The group $G$ is not $\D_5$, because
the line $x_3=x_4$ in the coordinate plane $(0,0,x_3,x_4,0)$ is invariant
for the permutation (2,5)(3,4); this prohibits a connection from $(0,0,a,0,0)$
to $(0,0,0,a,0)$. Therefore, $G=\Z_5$ and for $\dim P=2$ the lemma is proved.

\medskip
For $\Gamma=(\Z_2)^4\rtimes\Z_5$ or $\Gamma=(\Z_2)^3\rtimes\Z_5$,
a $\Gamma$-equivariant system can possess
homoclinic cycles with $\dim P=3$ or $\dim P=4$. A cycle with $\dim P=3$
involves equilibria on coordinate planes, e.g. $\xi=(a,b,0,0,0)$ and
its $\gamma$-symmetric copies, the connection from $\xi$ to $\gamma\xi$
belongs to the hyperplane $P=(d,e,f,0,0)$ and $\Sigma$ is either $\Z_2$
or $(\Z_2)^2$. A cycle with $\dim P=4$
involves equilibria on coordinate hyperplanes, e.g. $\xi=(a,b,c,0,0)$ and
its $\gamma$-symmetric copies, the connection from $\xi$ to $\gamma\xi$
belongs to the hyperplane $P=(d,e,f,g,0)$ and $\Sigma=\Z_2$.
The proof that $\Gamma$ is not any other subgroup of $\widetilde W$
is similar to the one presented for $\dim P=2$ and is omitted.
\qed

\bigskip
Theorem \ref{thmezi} and lemmas \ref{th9} and \ref{th8} enable us to find
a complete list of subgroups $\Gamma\subset$O(5) such that a $\Gamma$-equivariant
system (\ref{eq_ode})--(\ref{sym_ode}) can possess a simple homoclinic cycle $X$.
Theorem \ref{thmezi} states that any finite subgroup of O(5) is
either is one of those considered in lemmas \ref{th9} and
\ref{th8}, or it is a subgroup of O(4)$\times$O(1) or O(3)$\times$O(2).

In the latter cases the homoclinic cycle $X$ in $\R^5$ belongs
to a subspace $N\cong\R^3$ or $\R^4$. We denote by $\tilde X$ this
cycle in the restriction of the dynamical system into the respective subspace, by
$\tilde\Gamma$ the symmetry group of the restricted system,
by $\tilde P$ the fixed point subspace containing the homoclinic connection in $\tilde X$,
and by $\tilde\Sigma$ the isotropy subgroup of the subspace $\tilde P$.
By definition \ref{defn4} of the essential subgroup, $G_{\rm Ess}(X)=\tilde\Gamma$.
Type Z and type A' cycles are defined in terms of the isotypic decomposition
of $\tilde P^{\perp}$ under $\tilde\Sigma$. For cycles in $\R^3$ and
$\R^4$, the decomposition involves one or two isotypic components,
$$\tilde P^{\perp}=\tilde U_1\ \hbox{ or }\
\tilde P^{\perp}=\tilde U_1\oplus \tilde U_2.$$
If the decomposition has a single component, then the cycle is of type A',
otherwise it is of type Z. If the auxiliary subspace $\R^5\ominus N$ is $\R$,
then either it increases the dimension of $\tilde P$ or of an existing $\tilde U_r$, or it
constitutes a new isotypic component in the decomposition of $P^{\perp}$.
If $\R^5\ominus N=\R\oplus\R$, then for any $\R$ in the direct sum
the possibilities are the same.

For the cycles listed in table \ref{tab1}, $\dim\tilde P=2$. Note that
a system in $\R^4$, whose symmetry group is
$\tilde\Gamma=(\Z_2)^4\rtimes\Z_4$, can also possess a homoclinic cycle
for which $\dim\tilde P=3$ and $\tilde\Sigma=\Z_2$. No cycles exist with
$\dim\tilde P=3$ in a dynamical system that has any other symmetry group listed in this table.

Properties of homoclinic cycles are summarised in table \ref{tab2}.
We introduce classes of cycles (see the last column of table \ref{tab2})
in the next section in order to derive
conditions for asymptotic stability of the cycles (see table \ref{tab4}).
The first six lines in table \ref{tab2} describe genuinely \hbox{5-dimensional} homoclinic cycles
in dynamical systems, whose groups of symmetries are not
subgroups of O(4)$\times$O(1) and O(3)$\times$O(2); they are categorised
by lemma \ref{th8}.

\begin{table}[p]
\begin{center}
$$
\begin{array}{llllllll}
G_{\rm Ess} & \dim V_{\rm Ess} & N& \Gamma & \dim P &
{\rm d}(\Gamma)&{\rm d}(G) &\mbox{class}\\
\hline
(\Z_2)^5\rtimes\Z_5 & 5 & 5,10 & G_{\rm Ess}& 2 & 0 &0&\hbox{ 3-3-[2][3][1]}\\
(\Z_2)^5\rtimes\Z_5 & 5 & 5,10 & G_{\rm Ess}& 3 & 0 &0&\hbox{ 2-2-[2][1]}\\
(\Z_2)^5\rtimes\Z_5 & 5 & 5,10 & G_{\rm Ess}& 4 & 0 &0&\hbox{ 1-1}\\
(\Z_2)^4\rtimes\Z_5 & 5 & 10 & G_{\rm Ess}& 2 & 0 &0&\hbox{ 3-3-[2][3][1]}\\
(\Z_2)^4\rtimes\Z_5 & 5 & 10 & G_{\rm Ess}& 3 & 0 &0&\hbox{ 2-2-[2][1]}\\
(\Z_2)^4\rtimes\Z_5 & 5 & 10 & G_{\rm Ess}& 4 & 0 &0&\hbox{ 1-1}\\

(\Z_2)^4\rtimes\Z_4 & 4 & 4,8 & G_{\rm Ess} & 3 & 1 &1&\hbox{ 2-2-[2][1]}\\
(\Z_2)^4\rtimes\Z_4 & 4 & 4,8 & G_{\rm Ess} & 3 & 0 &0&\hbox{ 2-2-[2][1]}\\
(\Z_2)^4\rtimes\Z_4 & 4 & 4,8 & G_{\rm Ess}\times\Z_2 & 2 & 0 &0,1&\hbox{ 2-2-[2][1][3]}\\

(\Z_2)^3\rtimes\Z_4 & 4 & 8 & G_{\rm Ess} & 3 & 1 &1&\hbox{ 2-1-[12]}\\
(\Z_2)^3\rtimes\Z_4 & 4 & 8 & G_{\rm Ess} & 2 & 0 &0&\hbox{ 3-1-[123]}\\
(\Z_2)^3\rtimes\Z_4 & 4 & 8 & G_{\rm Ess} & 3 & 0 &0&\hbox{ 2-1-[12]}\\
(\Z_2)^3\rtimes\Z_4 & 4 & 8 & G_{\rm Ess}\times\Z_2 & 2 & 0 &0,1&\hbox{ 3-2-[12][3]}\\

\mO\rtimes\Z_4 & 4 & 12,24 & G_{\rm Ess} & 3 & 1 &1&\hbox{ 2-1-[12]}\\
\mO\rtimes\Z_4 & 4 & 12,24 & G_{\rm Ess} & 2 & 0 &0&\hbox{ 3-1-[123]}\\
\mO\rtimes\Z_4 & 4 & 12,24 & G_{\rm Ess} & 3 & 0 &0&\hbox{ 2-1-[12]}\\
\mO\rtimes\Z_4 & 4 & 12,24 & G_{\rm Ess}\times\Z_2 & 2 & 0 &0,1&\hbox{ 3-2-[12][3]}\\

\D_{2k}\rtimes\Z_k\rtimes\Z_2 & 4 & 4,2k &G_{\rm Ess} & 3 &1 &1&\hbox{ 2-1-[12]}\\
\D_{2k}\rtimes\Z_k\rtimes\Z_2 & 4 & 4,2k &G_{\rm Ess} & 2 &0 &0&\hbox{ 3-1-[123]}\\
\D_{2k}\rtimes\Z_k\rtimes\Z_2 & 4 & 4,2k &G_{\rm Ess} & 3 &0 &0&\hbox{ 2-1-[12]}\\
\D_{2k}\rtimes\Z_k\rtimes\Z_2 & 4 & 4,2k &G_{\rm Ess}\times\Z_2 & 2 &0 &0,1&\hbox{ 3-2-[12][3]}\\

(\Z_2)^4\rtimes\Z_4 & 4 &  4,8 & G_{\rm Ess} & 4 & 1 &1&\hbox{ 1-1}\\
(\Z_2)^4\rtimes\Z_4 & 4 &  4,8 & G_{\rm Ess} & 3 & 0 &0&\hbox{ 2-1-[12]}\\
(\Z_2)^4\rtimes\Z_4 & 4 &  4,8 & G_{\rm Ess} & 4 & 0 &0&\hbox{ 1-1}\\
(\Z_2)^4\rtimes\Z_4 & 4 &  4,8 & G_{\rm Ess}\times\Z_2 & 3 & 0 &0,1&\hbox{ 2-1-[1][2]}\\

(\Z_2)^3\rtimes\Z_3 & 3 & 3,6 & G_{\rm Ess} & 4 & 2 &2&\hbox{ 1-1}\\
(\Z_2)^3\rtimes\Z_3 & 3 & 3,6 & G_{\rm Ess} & 3 & 1 &1&\hbox{ 2-1-[12]}\\
(\Z_2)^3\rtimes\Z_3 & 3 & 3,6 & G_{\rm Ess}\times\Z_2 & 3 & 1 &1&\hbox{ 2-2-[1][2]}\\

(\Z_2)^3\rtimes\Z_3 & 3 & 3,6 & G_{\rm Ess} & 2 & 0 &0&\hbox{ 3-1-[123]}\\
(\Z_2)^3\rtimes\Z_3 & 3 & 3,6 & G_{\rm Ess} & 3 & 0 &0&\hbox{ 2-1-[12]}\\
(\Z_2)^3\rtimes\Z_3 & 3 & 3,6 & G_{\rm Ess}\times\Z_2 & 2 & 0 &0&\hbox{ 3-2-[12][3]}\\
(\Z_2)^3\rtimes\Z_3 & 3 & 3,6 & G_{\rm Ess}\times\Z_2 & 3 & 0 &1&\hbox{ 2-2-[1][2]}\\
(\Z_2)^3\rtimes\Z_3 & 3 & 3,6 & G_{\rm Ess}\times K & 2 & 0 &0,1,2&\hbox{ 3-2-[1][23]m}^*
\end{array}
$$
\end{center}
\caption{\label{tab2}
Simple homoclinic cycles, which can exist in a $\Gamma$-equivariant system
(\ref{eq_ode})--(\ref{sym_ode}) in $\R^5$. $N$ denotes the number of equilibria
involved in a cycle. d$(\Gamma)$ stands for $\dim\Fix(\Gamma)$ and d$(G)$
for $\dim\Fix(G_{\rm Ess})$. Label 3-2-[1][23]m$^*$ indicates
cycles which are either of class 3-2-[1][23] or 3-2-[1][23]m.}
\end{table}

\pagebreak
\begin{center}
$$
\begin{array}{llllllll}
G_{\rm Ess} & \dim V_{\rm Ess} & N& \Gamma & \dim P &
{\rm d}(\Gamma)&{\rm d}(G) &\mbox{class}\\
\hline
\D_4 & 2 & 2 & G_{\rm Ess} & 4 & 2 &2&\hbox{ 1-1}\\
\D_4 & 2 & 2 & G_{\rm Ess} & 3 & 1 &1&\hbox{ 2-1-[12]}\\
\D_4 & 2 & 2 & G_{\rm Ess} & 4 & 1 &1&\hbox{ 1-1}\\
\D_4 & 2 & 2 & G_{\rm Ess}\times\Z_2 & 3 & 1 &1,2&\hbox{ 2-2-[1][2]}\\

\D_4 & 2 & 2 & G_{\rm Ess} & 2 & 0 &0&\hbox{ 3-1-[123]}\\
\D_4 & 2 & 2 & G_{\rm Ess} & 4 & 0 &0&\hbox{ 1-1}\\
\D_4 & 2 & 2 & G_{\rm Ess} & 3 & 0 &0&\hbox{ 2-1-[12]}\\
\D_4 & 2 & 2 & G_{\rm Ess}\times\Z_2 & 2 & 0 &0&\hbox{ 3-2-[12][3]}\\
\D_4 & 2 & 2 & G_{\rm Ess}\times\Z_2 & 3 & 0,1 &0&\hbox{ 2-2-[1][2]}\\
\D_4 & 2 & 2 & G_{\rm Ess}\times\D_2 & 2 & 0 &0&\hbox{ 3-3-[1][3][2]}\\
\D_4 & 2 & 2 & G_{\rm Ess}\times K & 2 & 0 &0,1,2&\hbox{ 3-2-[1][23]m}^*
\end{array}
$$
\end{center}

\bigskip\noindent
\centerline{Continuation of table \ref{tab2}.}

\bigskip
Suppose now $\Gamma\subset$O(4)$\times$O(1), but $\Gamma\not\subset$O(3)$\times$O(2).
Then $X$ belongs to a subspace $N$ of $\R^5$ isomorphic to $\R^4$, as discussed above.
The group $\Gamma$ can be either $\tilde\Gamma$ or $\tilde\Gamma\times\Z_2$,\break
where $\tilde\Gamma$ is a symmetry group of a dynamical system in $\R^4$ and
$\Z_2$ acts on $\R$ as $-I$. The number of equilibria involved into $X$ is
the same as in the original cycle $\tilde X$,\break
$\dim\Fix(V_{\rm Ess}(X))=\dim\Fix(V_{\rm Ess}(\tilde X))$
and $G_{\rm Ess}(X)=\tilde\Gamma$. Other quantities shown in the table depend on
how $\tilde\Gamma$ acts on $\R=\R^5\ominus N$.

Table \ref{tab1} shows four $\tilde\Gamma$'s acting on $\R^4$ for which
homoclinic cycles can exist. For three of them, $\tilde\Sigma=\Z_2=<\sigma>$,
and for one $\tilde\Sigma=(\Z_2)^2$. If $\Gamma=\tilde\Gamma$ and it acts
trivially on $\R$, then $\dim\Fix({\Gamma})=1$, $\dim P=\dim\tilde P+1$,
$\Sigma=\tilde\Sigma$ and $X$ belongs to the same class as $\tilde X$.
If $\Gamma=\tilde\Gamma$ and it acts non-trivially on $\R$, or if
$\Gamma=\tilde\Gamma\times\Z_2$, then $\dim\Fix({\Gamma})=0$ and we have:
\begin{itemize}
\item[(i)] if $\Gamma=\tilde\Gamma$ and $\sigma$ acts on $\R$ as $-I$
(possible if $\tilde\Sigma=\Z_2$), then
$\dim P=\dim\tilde P$ and the cycle remains of type A';
\item[(ii)] if $\Gamma=\tilde\Gamma$, $\Sigma$ acts on $\R$ as $I$, $\gamma$ as
$-I$ (possible if $\gamma$ is of an even order), then \hbox{$\dim P=\dim\tilde P+1$}
and the cycle $X$ belongs to the same class as $\tilde X$;
\item[(iii)] if $\Gamma=\tilde\Gamma\times\Z_2$, then
$\dim P=\dim\tilde P$, $\Sigma=\tilde\Sigma\times\Z_2$ and the isotypic
decomposition of $P^{\perp}$ involves one component more, than the one
of $\tilde P^{\perp}$ (e.g. if $\tilde X$ belongs to the class 2-1-[12], then
$X$ belongs to 3-2-[12][3]).
\end{itemize}

Suppose now $\Gamma\subset$O(3)$\times$O(2).
Then $X$ belongs to a subspace of $\R^5$, $N$, that is isomorphic to $\R^3$.
The group $\Gamma$ can be either $\tilde\Gamma$, $\tilde\Gamma\times\Z_2$ or
$\tilde\Gamma\times K$, where $\tilde\Gamma$ is a symmetry group of
a dynamical system in $\R^3$, $\Z_2\subset$O(1) and $K\subset$O(2), $K\not\subset$O(1).
There are two distinct homoclinic cycles in $\R^3$,
for both of them $\tilde\Sigma=\Z_2=<\sigma>$ (see table \ref{tab1}).

If $\Gamma=\tilde\Gamma$ and it acts trivially on $\R^2=\R^5\ominus N$, then
$\dim\Fix({\Gamma})=2$,
$\dim P=\dim\tilde P+2$, $\Sigma=\tilde\Sigma$ and $X$ belongs to the 1-1 class.
If $\Gamma=\tilde\Gamma$ or $\Gamma=\tilde\Gamma\times\Z_2$ acting trivially
on one of $\R$ comprising $\R^2$, then depending on how $\Gamma$ acts on the remaining $\R$,
one of the cases (i)--(iii) takes place (however, in the cases (i) and (iii)
$\dim P=\dim\tilde P+1$, and in the case (ii) $\dim P=\dim\tilde P+2$).

If $\tilde\Gamma$ acts nontrivially in $\R^2$ or $\Gamma=\tilde\Gamma\times K$
(the two conditions are not mutually exclusive), then $\dim\Fix({\Gamma})=0$ and
exactly one of the following statements holds true:
\begin{itemize}
\item[(iv)] if $\Gamma=\tilde\Gamma$ and $\sigma$ acts on $\R^2$ as $-I$, then
$\dim P=2$ and the cycle $X$ belongs to the class 3-1-[123];
\item[(v)] if $\Gamma=\tilde\Gamma$, $\Sigma$ acts on $\R^2$ as $I$, $\gamma$ as
$-I$ (possible if $\gamma$ is of an even order), then $\dim P=4$ and the
cycle $X$ belongs to the class 1-1;
\item[(vi)] if $\Gamma=\tilde\Gamma$ and $\Gamma$ acts of $\R^2$ as $\D_k$
(only $k=2$, 3, 4 are possible), then $\dim P=3$ and the cycle $X$ belongs to the class 2-1-[12];
\item[(vii)] if $\Gamma=\tilde\Gamma\times\Z_2$, $\Z_2$ acts on one $\R$ as $I$
and on the other $R$ as $-I$, and $\sigma$ acts on $\R^2$ as $-I$, then
$\dim P=2$ and the cycle $X$ belongs to the class 3-2-[12][3];
\item[(viii)] if $\Gamma=\tilde\Gamma\times\Z_2$, $\Z_2$ acts on one $\R$ as $I$
and on the other $\R$ as $-I$, and $\sigma$ acts on one $\R$ as $I$ and
on the other as $-I$, then $\dim P=3$ and the cycle $X$ belongs to the class 2-2-[1][2];
\item[(ix)] if $\Gamma=\tilde\Gamma\times(\Z_2)^2$, $\Z_2$ acts on one $\R$ as $I$ and
on the other $\R$ as $-I$, and $\gamma$ permutes the two $\R$'s (possible if
$\gamma$ is of the second order), then $\dim P=2$ and the cycle belongs
to the class 3-3-[1][3][2];
\item[(x)] if $\Gamma=\tilde\Gamma\times K$, $K=\Z_r$ or $\D_r$ with $r>2$, then
$\dim P=2$ and the cycle belongs to a class 3-2-[1][23] or 3-2-[1][23]m,
which are discussed in appendix \ref{app_1}. Note that the conditions
for asymptotic stability of these cycles are identical.
\end{itemize}

\section{Classification of homoclinic cycles in $\R^n$}
\label{sec5}

By definition \ref{def5}, a homoclinic cycle is
structurally stable, if there exists a subgroup $\Sigma\subset\Gamma$ such that
the connection from $\xi$ to $\gamma\xi$ belongs to a fixed-point subspace
$P=\Fix(\Sigma)$. We introduce a classification of simple homoclinic cycles
in $\R^n$ that is based on the actions
of $\Sigma$ and $\gamma$ (note that $\gamma\notin\Sigma$) on $P^{\perp}$.
The classification suffices to determine conditions for stability
of homoclinic cycles in $\R^5$ --- for each class, they have the form of
inequalities for eigenvalues of the linearisation $df(\xi)$ (see table \ref{tab4}).

Let ${\bf e}_k$, $1\le k\le K$, be a basis in $P^{\perp}$ comprised of
eigenvectors of $df(\xi)$, and let ${\bf h}_k$, $1\le k\le K$, be a basis
in $P^{\perp}$ comprised of eigenvectors of $df(\gamma\xi)$. We assume that
${\bf e}_1$ is the contracting eigenvector of $df(\xi)$, ${\bf h}_1$ is
the expanding eigenvector of $df(\gamma\xi)$, and the remaining transverse
eigenvectors are related: ${\bf h}_k=\gamma{\bf e}_k$, $2\le k\le K$. Suppose
\begin{equation}\label{isdec}
P^{\perp}=U_1\oplus\ldots\oplus U_J
\end{equation}
is the isotypic decomposition of $P^{\perp}$ under the action of $\Sigma$. Let
the eigenvectors in the basis be ordered in such a way that the ${\bf h}_k$,
belonging to one isotypic component, $U_j$, (whose dimension we denote by
$l_j$), have consecutive indices, $k=s+1,\ldots,s+l_j$; namely, $U_1$ is spanned
by ${\bf h}_1,\ldots,{\bf h}_{l_1}$,
$U_2$ by ${\bf h}_{l_1+1},\ldots,{\bf h}_{l_1+l_2}$, etc. Each isotypic
component $U_j$ is also spanned by the eigenvectors ${\bf e}_k$ for some
$k=i_{s_{j-1}+1},i_{s_{j-1}+2},\ldots,i_{s_j}$, where we have denoted
$s_j=l_1+\ldots+l_j$. We label such a homoclinic cycle by a sequence
of numbers, where subsequences associated with individual isotypic components
are enclosed in square brackets:
$$[i_1,i_2,\ldots,i_{l_1}][i_{l_1+1},i_{l_1+2},\ldots,i_{s_2}]\ldots
[i_{s_{J-1}+1},i_{s_{J-1}+2},\ldots,i_{s_J}].$$

However, not all possible permutations and combinations of brackets is
encountered in a homoclinic cycle, as shown in the following lemma.

\begin{lemma}\label{lemn3}
Let $X$ be a simple robust homoclinic cycle in a system (\ref{eq_ode})--(\ref{sym_ode}).
Consider the isotypic decomposition of $P^{\perp}$ under the action of $\Sigma$,
\begin{equation}\label{isdec1}
P^{\perp}=U_1\oplus\ldots\oplus U_J.
\end{equation}
Recall that for our ordering of eigenvectors ${\bf h}_1\in U_1$. Denote by
$U$ the isotypic component in (\ref{isdec1}) that contains ${\bf e}_1$. Then\\
(i) $\dim U_1=\dim U$.\\
(ii) $\gamma{\bf e}_j\notin U_1$ for any ${\bf e}_j\notin U$.
\end{lemma}

\proof
(i) Denote $S=\Sigma\cap\gamma\Sigma\gamma^{-1}$ and $Q=P^{\perp}\cap\gamma P^{\perp}$.
Consider the isotypic decomposition of the subspace $Q$ under the action of $S$,
$$Q=V_1\oplus\ldots\oplus V_I,$$
where $V_1$ is the isotypic component in which $S$ acts trivially. The isotypic
decomposition of $\gamma P^{\perp}$ under the action of $\gamma\Sigma\gamma^{-1}$ is
$$\gamma P^{\perp}=\gamma U_1\oplus\ldots\oplus \gamma U_J.$$
By definition of simple homoclinic cycles, $P^{\perp}=Q\oplus<{\bf h}_1>$ and
$\gamma P^{\perp}=Q\oplus<\gamma{\bf e}_1>$. The group $S$ acts trivially on
$<{\bf h}_1>$ and $<\gamma{\bf e}_1>$. Therefore, $U_1=V_1\oplus<{\bf h}_1>$
and $\gamma U=V_1\oplus<\gamma{\bf e}_1>$; this implies
$\dim U_1=\dim V_1+1=\dim U$.

(ii) As we have found, $U=\gamma^{-1}V_1\oplus<{\bf e}_1>$. Therefore,
the condition of the lemma implies ${\bf e}_j\notin\gamma^{-1}V_1$, and hence
$\gamma{\bf e}_j\notin V_1$. Since $\gamma{\bf e}_j\in Q$ and $Q\perp<{\bf h}_1>$,
$\gamma{\bf e}_j\notin V_1\oplus<{\bf h}_1>=U_1$.
\qed

All classes of homoclinic cycles in $\R^5$ are listed in table \ref{tab3}. The sequences
defined above are supplemented by two numbers: the dimension of $P^{\perp}$ and
the number of isotypic components (e.g. 3-1-[123] labels a cycle with
a three-dimensional $P^{\perp}$ comprised of a single isotypic component).

\section{Poincar\'e maps for homoclinic cycles in $\R^5$}
\label{sec6}

Following \cite{ks94,km04,pa11}, in order to examine stability of a homoclinic
cycle we consider a Poincar\'e map near the cycle. In subsection \ref{hetc}
we have defined radial, contracting, expanding and transverse eigenvalues
of the linearisation $df(\xi)$. Let $(\tilde{\bf u},\tilde v,\tilde w,\tilde{\bf z})$
be coordinates in the coordinate system with the origin at $\xi$ and the basis
comprised of the associated eigenvectors in the following order: radial,
contracting, expanding and transverse.

If $\tilde\delta$ is small, in a $\tilde\delta$-neighbourhood of $\xi$
the system (\ref{eq_ode}) can be approximated by the linear system\footnote
{We assume here that all eigenvalues are real. The system under consideration
can have a pair of complex conjugate radial eigenvalues, if the dimension
of the radial eigenspace is larger than one, or it can have a pair of complex
conjugate transverse eigenvalues, if the homoclinic cycle is of the classes
3-1-[123] or 3-2-[1][23]. The radial eigenvalues are not relevant in the study
of stability. If transverse eigenvalues are complex, the estimates
$$k_1(|z_1|+|z_2|)|w|^{-t/e}\le|z_1|\le K_1(|z_1|+|z_2|)|w|^{-t/e},\
k_2(|z_1|+|z_2|)|w|^{-t/e}\le|z_2|\le K_2(|z_1|+|z_2|)|w|^{-t/e},$$
where $t={\rm Re}(t_1)={\rm Re}(t_2)$, can be employed in the proofs
of stability and instability, respectively, instead of the exact expressions
$$z_1=z_1|w|^{-t_1/e},\ z_2=z_2|w|^{-t_2/e}.$$
Since only the values of exponents are important in the proofs,
in the conditions for stability (see table \ref{tab4}) $t_i$ is replaced by
Re$(t_i)$, and no other modifications are required.}
\begin{equation}\label{lmap}
\begin{array}{l}
\dot u_l=-r_lu_l,\ 1\le l\le n_r\\
\dot v=-c v\\
\dot w=e w\\
\dot z_l=t_lz_l,\ 1\le l\le n_t.
\end{array}
\end{equation}
Here, $({\bf u},v,w,{\bf z})$ denote the scaled coordinates
$({\bf u},v,w,{\bf z})=
(\tilde{\bf u},\tilde v,\tilde w,\tilde{\bf z})/\tilde\delta$.

Let $({\bf u}_0,v_0)$ be the point in $\gamma^{-1}P$ where the trajectory
$\gamma^{-1}\kappa$ intersects with the sphere $|{\bf u}|^2+v^2=1$, and $\bf q$
be local coordinates in the hyperplane tangent to the sphere at the point
$({\bf u}_0,v_0)$. We consider two crossections:
$$
\widetilde H^{(out)}=\{({\bf u},v,w,{\bf z})~:~|{\bf u}|,|v|,|{\bf z}|\le1, w=1\}
$$
and
$$
\widetilde H^{(in)}=\{({\bf q},w,{\bf z})~:~|{\bf q}|,|w|,|{\bf z}|\le1\}.
$$

Near $\xi$, trajectories of system (\ref{eq_ode})
can be approximated by a local map (called the {\em first return map})
$\tilde\phi:\widetilde H^{(in)}\to\widetilde H^{(out)}$ that associates a point, where
a trajectory crosses $\widetilde H^{(out)}$ with the point, where the trajectory
crossed $\widetilde H^{(in)}$. The global map
$\tilde\psi:\widetilde H^{(out)}\to\gamma\widetilde H^{(in)}$ associates a point where
a trajectory crosses $\gamma\widetilde H^{(in)}$ with the point where its previously
crossed $\widetilde H^{(out)}$. The Poincar\'e map is the superposition
$\widetilde g=\tilde\psi\tilde\phi$. The $w$- and $\bf z$-components of the map
$\widetilde g$ are independent of $\bf q$: this was shown in \cite{pa11,op12}
for slightly different systems, but the proof can be trivially modified to serve
the case considered here. Thus, one can define the map $g(w,{\bf z})$ that is
the restriction of the map $\widetilde g$ into the $(w,{\bf z})$ subspace.
The stability properties of fixed points of the maps $\widetilde g$ and $g$ are
identical; hence, stability of a cycle is determined by stability of the fixed
point $(w,{\bf z})=\bf 0$ of the map $g$.

Denote by $\phi$ and $\psi$ the restrictions of $\tilde\psi$ and $\tilde\phi$
into $P^{\perp}$. In the leading order, the map $\phi$~is
\begin{equation}\label{phmap}
\phi(w,\{z_s\})=(v_0w^{c/e},\{z_s|w|^{-t_s/e}\})
\end{equation}
(we use the coordinates $(w,{\bf z})$ in $H^{(in)}$ and $(v,{\bf z})$
in $H^{(out)}$). Note that the local map is expressed by (\ref{phmap})
for any homoclinic cycle of whichever class.

Expressions for global maps are different for different classes
of homoclinic cycles. The conditions of lemma \ref{lem0}
(see appendix \ref{app_1}) are satisfied for all simple homoclinic cycles in
$\R^5$ except for the 3-2-[1][23] cycles. We do not
consider the 3-2-[1][23] cycles henceforth in this section; the global map
for them is derived in appendix \ref{app_1}. By the lemma, each
isotypic component of $P^{\perp}$ is an invariant subspace of the map $\psi$.

Generically, in the leading order the global map $\psi$ is linear in each
isotypic component. The matrix, $C$, of the linear map
\begin{equation}\label{psmap}
(w^{\gamma\xi},{\bf z}^{\gamma\xi})=\psi(v^{\xi},{\bf z}^{\xi})=C
\left(
\begin{array}{c}
v^{\xi}\\
{\bf z}^{\xi}
\end{array}
\right),
\end{equation}
(here superscripts indicate whether the components are in the basis
of eigenvectors of $df(\xi)$ or of $df(\gamma\xi)$) is the product $C=BA$,
where $A$ is the matrix of the map $\psi$ in the basis of eigenvectors of
$df(\xi)$, and $B$ is the matrix of transformation of $(v^{\xi},{\bf z}^{\xi})$
into $(w^{\gamma\xi},{\bf z}^{\gamma\xi})$. In the study of stability, we focus
on the location of blocks in $C=\{c_{ij}\}$ that vanish because the map $\psi$
has invariant subspaces. Generically, $c_{ij}\ne0$, if ${\bf e}_i$ and
${\bf h}_j$ belong to the same isotypic component in the decomposition
(\ref{isdec}). Hence, the location of zero entries of $C$ can be determined
applying the classification presented in section \ref{sec5}. Below we list
exhaustively the possible forms of matrices $C$ (where non-zero entries are
shown by $*$) for various types of homoclinic cycles in $\R^5$, and determine
the general forms of Poincar\'e maps (see table \ref{tab3}).

\medskip
If the subspace $P^{\perp}$ is one-dimensional, then $C$ is
an $1\times1$ matrix and the Poincar\'e map is just $g(w)=c_{11}w^{c/e}$.

\medskip
If the subspace $P^{\perp}$ is two-dimensional, then the classification of
simple homoclinic cycles in $\R^4$ is applicable, since for such cycles
$P^{\perp}$ is two-dimensional. Alternatively, note that such cycles are either
of type A$'$ (if the decomposition of $P^{\perp}$ under
$\Sigma$ has only one isotypic component) or Z (if there are two components).
In the former case the cycle is classified as 2-1-[12], generically none
entries of its matrix $C$ vanish, and thus the Poincar\'e map is
$g(w,z)=(c_{11}w^{c/e}+c_{12}z|w|^{-t/e},c_{21}w^{c/e}+c_{22}z|w|^{-t/e})$.
In the latter case the cycles are of the 2-2-[1][2] or 2-2-[2][1] classes,
the matrices of the global map $\psi$ are
\begin{equation}\label{type22}
\begin{array}{cc}
\left(
\begin{array}{cc}
*&0\\
0&*
\end{array}
\right)&
\left(
\begin{array}{cc}
0&*\\
*&0
\end{array}
\right)\\
\\
\hbox{ 2-2-[1][2] }&\hbox{ 2-2-[2][1] }
\end{array}
\end{equation}
and the Poincar\'e maps are
$g(w,z)=(c_{11}w^{c/e},c_{22}z|w|^{-t/e})$ (for the 2-2-[1][2] cycle) and
$g(w,z)=(c_{12}z|w|^{-t/e},c_{21}w^{c/e})$ (for the 2-2-[2][1] cycle).

\medskip
If the subspace $P^{\perp}$ is three-dimensional, then homoclinic cycles
are of types A$'$ (if the decomposition (\ref{isdec}) involves just one isotypic
component), Z (if the decomposition involves three components), or of other
types not studied so far. For a type A$'$ cycle listed in tables 1 and 2
as 3-1-[123], generically all entries of $C$ are non-zero. For type Z
cycles, $A$ is a diagonal matrix
and $B$ is a permutation matrix (provided vectors ${\bf e}_j$ and ${\bf h}_j$
in the bases are normalised), and hence $C$ has one of the following forms:
\begin{equation}\label{cyc33}
\begin{array}{cccc}
\left(
\begin{array}{ccc}
*&0&0\\
0&*&0\\
0&0&*
\end{array}
\right)&
\left(
\begin{array}{ccc}
*&0&0\\
0&0&*\\
0&*&0
\end{array}
\right)&
\left(
\begin{array}{ccc}
0&*&0\\
*&0&0\\
0&0&*
\end{array}
\right)&
\left(
\begin{array}{ccc}
0&*&0\\
0&0&*\\
*&0&0
\end{array}
\right)\\
\\
\hbox{ 3-3-[1][2][3] }&\hbox{ 3-3-[1][3][2] }&\hbox{ 3-3-[2][1][3] }&\hbox{ 3-3-[2][3][1] }
\end{array}
\end{equation}

Now suppose decomposition (\ref{isdec}) involves two isotypic components. Then
the cycles are either 3-2-[$\cdot][\cdot\cdot$], or 3-2-[$\cdot\cdot][\cdot$],
where dots stand for a permutation of indices $\{1,2,3\}$. Five permutations
are possible: [1][23], [2][13], [12][3], [13][2] and [23][1]. However,
by lemma \ref{lemn3} only two of them are realised in homoclinic cycles.
The matrices of global maps for these homoclinic cycles have the following forms:
\begin{equation}\label{cyc32}
\begin{array}{cc}
\left(
\begin{array}{ccc}
*&0&0\\
0&*&*\\
0&*&*
\end{array}
\right)&
\left(
\begin{array}{ccc}
*&*&0\\
*&*&0\\
0&0&*
\end{array}
\right)\\
\\
\hbox{ 3-2-[1][23] }&\hbox{ 3-2-[12][3] }
\end{array}
\end{equation}
Poincar\'e maps for these cycles are listed in table \ref{tab3}. The exponents $a_j$,
$j=$1,2,3, can be expressed in terms of eigenvalues of the linearisation
$df(\xi)$ by the relations
\begin{equation}\label{exeig}
a_1=c/e,\ a_2=-t_1/e\hbox{ (or $-t/e$ if dim}P^{\perp}=1),\ a_3=-t_2/e.
\end{equation}

\begin{table}[t]
\begin{center}
$$
\begin{array}{lll}
\mbox{Class} & \mbox{Poincar\'e map} & \mbox{Type}\\
\hline
\hbox{1-1-[1]} & g(w)=(c_{11}w^{a_1}) & {\rm A',Z}\\
\hline
\hbox{2-1-[12]} & g(w,z)=(c_{11}w^{a_1}+c_{12}z|w|^{a_2},
c_{21}w^{a_1}+c_{22}z|w|^{a_2})&{\rm A'}\\
\hbox{2-2-[1][2]} & g(w,z)=(c_{11}w^{a_1},c_{22}z|w|^{a_2}) & {\rm Z}\\
\hbox{2-2-[2][1]} & g(w,z)=(c_{12}z|w|^{a_2},c_{21}w^{a_1}) & {\rm Z}\\
\hline
\hbox{3-1-[123]} & g(w,z_1,z_2)=(c_{11}w^{a_1}+c_{12}z_1|w|^{a_2}+c_{13}z_2|w|^{a_3},
&{\rm A'}\\
& c_{21}w^{a_1}+c_{22}z_1|w|^{a_2}+c_{23}z_2|w|^{a_3},
c_{31}w^{a_1}+c_{32}z_1|w|^{a_2}+c_{33}z_2|w|^{a_3})&\\
\hline
\hbox{3-2-[12][3]} & g(w,z_1,z_2)=(c_{11}w^{a_1}+c_{12}z_1|w|^{a_2},
c_{21}w^{a_1}+c_{22}z_1|w|^{a_2},c_{33}z_2|w|^{a_3})& \\
\hbox{3-2-[1][23]} & g(w,z_1,z_2)=(c_{11}w^{a_1},
c_{22}z_1|w|^{a_2}+c_{23}z_2|w|^{a_3},c_{32}z_1|w|^{a_2}+c_{33}z_2|w|^{a_3})& \\
\hbox{3-2-[1][23]m} & g(w,z_1,z_2)=(c_{11}w^{a_1}+|w|^{a_2k}{\rm Re}
((c_{12}+\ri c_{13})(z_1+\ri z_2)^k),\\
&\hspace*{1cm}c_{22}z_1|w|^{a_2}+c_{23}z_2|w|^{a_2},
-c_{23}z_1|w|^{a_2}+c_{22}z_2|w|^{a_2})& \\
\hline
\hbox{3-3-[1][2][3]} & g(w,z_1,z_2)=
(c_{11}w^{a_1},c_{22}z_1|w|^{a_2},c_{33}z_2|w|^{a_3})& {\rm Z}\\
\hbox{3-3-[1][3][2]} & g(w,z_1,z_2)=(c_{11}w^{a_1},
c_{23}z_2|w|^{a_3},c_{32}z_1|w|^{a_2})& {\rm Z}\\
\hbox{3-3-[2][1][3]} & g(w,z_1,z_2)=(c_{12}z_1|w|^{a_2},
c_{21}w^{a_1},c_{33}z_2|w|^{a_3})& {\rm Z}\\
\hbox{3-3-[2][3][1]} & g(w,z_1,z_2)=(c_{12}z_1|w|^{a_2},
c_{23}z_2|w|^{a_3},c_{31}w^{a_1})& {\rm Z}
\end{array}
$$
\end{center}
\caption{\label{tab3}
Poincar\'e maps for different classes of homoclinic cycles in $\R^5$. Last
column indicates cycles of types A$'$ or Z. $a_j$ are the ratios (\ref{exeig})
of eigenvalues of the linearisation.}
\end{table}

\section{Stability}\label{scyc}

\subsection{Stability of cycles of types A and Z}\label{stabAZ}

In this subsection we review conditions for asymptotic stability of cycles
of types A \cite{km95a,km95b} and Z \cite{op12}. Following the long established
tradition, we use the expression ``asymptotic stability of homoclinic cycles'',
while in fact we speak about asymptotic stability of a homoclinic network.
Note that a homoclinic cycle can never be asymptotically stable, as discussed
in section 2.5 of \cite{op12}, while the respective homoclinic network can.

\begin{theorem}\label{th_1}(\cite{km95a,km95b}, adapted for homoclinic
cycles). Let $-c$, $e$ and $t_j$, $1\le j\le J$,
be the contracting, expanding and transverse eigenvalues of $df(\xi)$
for the steady states $\xi$ involved in a homoclinic cycle of type A.
\begin{itemize}
\item[(a)] If $c>e$ and $t_j<0$ for all $1\le j\le J$, then the cycle is
asymptotically stable.
\item[(b)] If $c<e$ or $t_j>0$ for some $j$ then the cycle is completely
unstable.
\end{itemize}
\end{theorem}

Stability of the fixed point $(w,{\bf z})=\bf 0$ of the map $g$ associated
with a type Z homoclinic cycle was studied in \cite{op12} by considering
the map in the coordinates
\begin{equation}\label{newc}
\mbe=(\ln|w|,\ln|z_1|,...,\ln|z_{n_t}|),
\end{equation}
in which the map is linear:
\begin{equation}\label{fmap}
g\mbe=M\mbe+{\bf F}.
\end{equation}
Here
\begin{equation}\label{esm}
M=B\left(
\begin{array}{ccccc}
a_1&0&0&\ldots&0\\
a_2&1&0&\ldots&0\\
a_3&0&1&\ldots&0\\
.&.&.&\ldots&.\\
a_N&0&0&\ldots&1
\end{array}
\right)
\end{equation}
is the transition matrix of the map $g$, $B$ is a permutation matrix (see
section \ref{sec6}), and the entries $a_j$ in the matrix of the local map are
\begin{equation}\label{coeB}
a_1=c/e\mbox{ and }a_{j+1}=-t_j/e,\ 1\le j\le J.
\end{equation}

Any permutation is a superposition of cyclic permutations. We assume that
vectors ${\bf e}_j$ in the basis are ordered in such a way that the first $n_s$
vectors (the subscript $s$ stands for significant) are involved in a single
cyclic permutation and ${\bf e}_1$ is the contracting eigenvector of $df(\xi)$.
Matrix $B$ permutes the vectors:
${\bf e}_1\to{\bf e}_2\to\ldots\to{\bf e}_{n_s}\to{\bf e}_1$.
Any eigenvalue of the upper left $n_s\times n_s$ submatrix of $M$ is also
an eigenvalue of $M$, because the upper right $(J-n_s)\times n_s$ submatrix
of $M$ vanishes. These eigenvalues are called significant; generically
they differ from one in absolute value. All other eigenvalues
of $M$ are one in absolute value. Denote by $\lambda_{\max}$ the largest in
absolute value significant eigenvalue of $M$ and by ${\bf v}^{\max}$ the
associated eigenvector. Conditions for asymptotic and fragmentary
asymptotic stability of type Z cycles in terms of $\lambda_{\max}$ and
components of ${\bf v}^{\max}$ are stated in theorems \ref{th_2} and \ref{th_3}:

\begin{theorem}\cite{op12}\label{th_2}
Let $M$ be the transition matrix of a homoclinic cycle of type Z. Suppose
all transverse eigenvalues of $df(\xi)$ are negative.
\begin{itemize}
\item[(a)] If $|\lambda_{\max}|>1$, then the cycle is asymptotically stable.
\item[(b)] If $|\lambda_{\max}|<1$, then the cycle is completely unstable.
\end{itemize}
\end{theorem}

\begin{theorem}\cite{op12}\label{th_3}
Let $M$ be the transition matrix of a homoclinic cycle of type Z.
The cycle is fragmentarily asymptotically stable if and only if
the following conditions are satisfied:
\begin{itemize}
\item[(i)] $\lambda_{\max}$ is real;
\item[(ii)] $\lambda_{\max}>1$;
\item[(iii)] $v_l^{\max}v_q^{\max}>0$ for all $l$ and $q$, $1\le l,q\le N$.
\end{itemize}
\end{theorem}

\subsection{Stability of homoclinic cycles in $\R^5$}\label{stab5}

Conditions for asymptotic stability and fragmentary asymptotic stability
for various classes of homoclinic cycles are presented in table \ref{tab4}. For
type A$'$ cycles the conditions follow from theorem \ref{th_1}.

\medskip
For type Z cycles the conditions are determined from theorems \ref{th_2} and
\ref{th_3} by calculating eigenvalues and eigenvectors of transition matrices.
For the 2-2-[$\cdot][\cdot$] cycles the transition matrices are
\begin{equation}\label{tr22}
\begin{array}{cc}
\left(
\begin{array}{cc}
a_1&0\\
a_2&1
\end{array}
\right)&
\left(
\begin{array}{ccc}
a_2&1\\
a_1&0
\end{array}
\right)\\
\\
\hbox{ 2-2-[1][2] }&\hbox{ 2-2-[2][1]}
\end{array}
\end{equation}
The first and second matrices have a one- and two-dimensional significant
subspace, respectively. Calculating the eigenvectors and eigenvalues, we
determine the conditions for asymptotic stability listed in table \ref{tab4}
(previously found in \cite{km04,op12,pa11}). When such a cycle is not
asymptotically stable, it is completely unstable.

\begin{table}[t]
\begin{center}
$$
\begin{array}{ll}
\mbox{Class} & \mbox{Conditions for stability} \\
\hline
\hbox{1-1-[1]} & \mbox{A. s.: } c>e\\
\hline
\hbox{2-1-[12]} & \mbox{A. s.: } c>e,\ t<0\\
\hline
\hbox{2-2-[1][2]} & \mbox{A. s.: } c>e,\ t<0\\
\hbox{2-2-[2][1]} & \mbox{A. s.: } c-t>e,\ t<0\\
\hline
\hbox{3-1-[123]} & \mbox{A. s.: } c>e,\ t_1<0,\ t_2<0\\
\hline
\hbox{3-2-[12][3]} & \mbox{A. s.: } c>e,\ t_1<0,\ t_2<0\\
\hbox{3-2-[1][23]}, & \mbox{A. s.: } c>e,\ t_1<0,\ t_2<0\\
\hbox{3-2-[1][23]m} & \mbox{A. s.: } c>e,\ t_1<0,\ t_2<0\\
\hline
\hbox{3-3-[1][2][3]} & \mbox{A. s.: } c>e,\ t_1<0,\ t_2<0\\
\hbox{3-3-[1][3][2]} & \mbox{A. s.: } c>e,\ t_1<0,\ t_2<0\\
            & \mbox{F. a. s.: } c>e,\ ct_1+et_2<0,\ ct_2+et_1<0\\
\hbox{3-3-[2][1][3]} & \mbox{A. s.: } c-t_1>e,\ t_1<0,\ t_2<0\\
\hbox{3-3-[2][3][1]} & \mbox{A. s.: } c-t_1-t_2>e,\ t_1<0,\ t_2<0\\
            & \mbox{F. a. s.: } c-t_1-t_2>e,\ t_1t_2+ce>0,\ ct_1^3+et_2^3<0
\end{array}
$$\end{center}
\caption{\label{tab4}
Conditions for asymptotic stability and fragmentary asymptotic stability of
different classes of homoclinic cycles in $\R^5$ in terms of eigenvalues of
the linearisation $df(\xi)$.}
\end{table}

The transition matrices of the 3-3-[$\cdot][\cdot][\cdot$] cycles are
\begin{equation}\label{tr33}
\begin{array}{cccc}
\left(
\begin{array}{ccc}
a_1&0&0\\
a_2&1&0\\
a_3&0&1
\end{array}
\right)&
\left(
\begin{array}{ccc}
a_1&0&0\\
a_3&0&1\\
a_2&1&0
\end{array}
\right)&
\left(
\begin{array}{ccc}
a_2&1&0\\
a_1&0&0\\
a_3&0&1
\end{array}
\right)&
\left(
\begin{array}{ccc}
a_2&1&0\\
a_3&0&1\\
a_1&0&0
\end{array}
\right)\\
\\
\hbox{ 3-3-[1][2][3] }&\hbox{ 3-3-[1][3][2] }&\hbox{ 3-3-[2][1][3] }&\hbox{ 3-3-[2][3][1] }
\end{array}
\end{equation}
The dimension of the significant subspace of the first, second, third and
fourth matrix is one, one, two and three, respectively. For the matrices
with one- and two-dimensional significant subspaces, the conditions of theorems
\ref{th_2} and \ref{th_3} can be expressed in terms of $a_i$ by explicitly
calculating the eigenvectors; the respective maps are either asymptotically
stable or completely unstable. For the fourth matrix, the relations between
the entries $a_i$, that are equivalent to the conditions of the theorems, are
derived in appendix \ref{app_2}. Substituting (\ref{coeB}), we obtain
the conditions listed in table \ref{tab4}.

\medskip
The Poincar\'e map for the 3-2-[12][3] cycle is
$$g(w,z_1,z_2)=(c_{11}w^{a_1}+c_{12}z_1|w|^{a_2},
c_{21}w^{a_1}+c_{22}z_1|w|^{a_2},c_{33}z_2|w|^{a_3}),$$
implying that the condition $a_3>0$ is necessary for fragmentary stability of
the cycle. The first two components of $g$ do not dependent $z_2$, and for them
we use the conditions for asymptotic stability of the 2-1-[12] cycle.

The Poincar\'e map for the 3-2-[1][23] cycle is
$$g(w,z_1,z_2)=(c_{11}w^{a_1},
c_{22}z_1|w|^{a_2}+c_{23}z_2|w|^{a_3},c_{32}z_1|w|^{a_2}+c_{33}z_2|w|^{a_3}).$$
Using this expression one can easily derive the necessary and sufficient
conditions for asymptotic stability: $a_1>1$, $a_2>0$ and $a_3>0$.
The asymptotic stability of the 3-2-[1][23]m cycle is studied in appendix \ref{app_3}.

\section{Discussion}\label{sec_conc}

We have found in the present paper all simple homoclinic cycles in $\R^5$ and
the respective conditions for asymptotic stability. Perhaps, the most
fascinating finding is that no new kinds of homoclinic cycles in $\R^5$ are
revealed. They are either of type Z studied in \cite{op12}, or belong
to a subspace of $\R^5$ isomorphic to $\R^3$ or $\R^4$. A question arises,
whether homoclinic cycles of other types exist in $\R^n$ for $n>5$.

The conditions for stability of type Z are derived in \cite{op12}.
In $\R^5$, only cycles of type Z can be fragmentarily asymptotically stable;
other cycles can be asymptotically stable or completely unstable. A cycle, that
is not of type Z, is asymptotically stable if and only if the contracting
eigenvalue is larger than the expanding and all transverse eigenvalues are
negative. Is this simple criterion for asymptotic stability of cycles that
are not of type Z remains valid in $\R^n$ for $n>5$ is an open question.

A natural continuation of the present work is an investigation of resonance
bifurcations of simple homoclinic cycles in $\R^5$, similar to the study
\cite{dh09} for homoclinic cycles in $\R^4$. Another possible continuation
is an investigation of simple heteroclinic cycles in $\R^4$ using
the homomorphism $\mH\times\mH\to$SO(4)
or of simple heteroclinic cycles in $\R^5$ using theorem \ref{thmezi}.

\subsection*{Acknowledgements}

My research was financed in part by the grant 11-05-00167-a from
the Russian foundation for basic research. Several visits to the
Observatoire de la C\^ote d'Azur (France) were
supported by the French Ministry of Higher Education and Research.

\appendix
\section{Global map for the 3-2-[1][23] homoclinic cycles}\label{app_1}

Consider a map $\psi:\R^3\to\R^3$ equivariant under a symmetry group
$\Sigma\subset$O(3). Suppose
\begin{itemize}
\item[(a)] $\Sigma$ decomposes $\R^3$ into two isotypic components
$$\R^3=U_1\oplus U_2,$$
where the dimension of $U_1$ is one and of $U_2$ is two;
\item[(b)] for any ${\bf x}\in\R^3$ there exists $\sigma\in\Sigma$ such that
$\sigma{\bf x}\ne\bf x$.
\end{itemize}
Consider the $(x,z)$ coordinates in $\R^3$, where $x$ is the coordinate in $U_1$
and $z$ in $U_2$. In this appendix we determine the leading terms of the
expansion of $\psi$ in small $x$ and $z$. We use the following lemma \cite{op12}:

\begin{lemma}\label{lem0}
Let a group $\Sigma$ act on a linear space $V$. Consider the isotypic
decomposition of $V$ under the action of $\Sigma$:
$$V=U_0\oplus U_1\oplus\ldots\oplus U_K.$$
Suppose
\begin{itemize}
\item the action of $\Sigma$ on $U_0$ is trivial;
\item any $\sigma\in\Sigma$ acts on a $U_k$, $1\le k\le K$, either as $I$ or
as $-I$.
\end{itemize}
Then for any collection of subscripts $1\le i_1,\ldots,i_l\le K$ there exists
a subgroup $G_{i_1,\ldots,i_l}\subset\Sigma$ such that the subspace
$$V_{i_1,\ldots,i_l}=U_0\oplus U_{i_1}\oplus\ldots\oplus U_{i_l}$$
is a fixed point subspace of the group $G_{i_1,\ldots,i_l}$.
\end{lemma}

By (a) and (b), there exists a symmetry $\sigma_1\in\Sigma$ such that
$\sigma_1(x,0)=(-x,0)$. $\Sigma$ can act on $U_2$ in three ways:
\begin{itemize}
\item[(i)] there exists $\sigma_2\in\Sigma$ such that $\sigma_2\ne\sigma_1^k$
for any $k$ and $\sigma_2(0,z)=(0,-z)$;
\item [(ii)] no $\sigma_2$ satisfying (i) exists; there exists
$\sigma_3\in\Sigma$ such that $\sigma_3(0,z)=(0,{\rm e}^{2\pi{\rm i}/k}z)$,
where $k>1$ is odd, and $\sigma_1(x,z)=(-x,z)$;
\item [(iii)] no $\sigma_2$ satisfying (i) exists; there exist
$\sigma_3\in\Sigma$ such that $\sigma_3(0,z)=(0,{\rm e}^{2\pi{\rm i}/k}z)$,
where $k>1$ is odd, and $\sigma_1(x,z)=(-x,-z)$.
\end{itemize}

Lemma 1 implies that in case (i) the $x-$ and $z$-components of $\psi$ are
$$\psi^x=xF(x,z,\bar z),\quad\psi^z=zG(x,z,\bar z)+\bar z^sH(x,z,\bar z),$$
where $F$ is real, generically $F(0,0,0)\ne0$ and $G(0,0,0)\ne0$, $s>0$ is odd
(it is determined by $\Sigma$). In cases (ii) and (iii) the components
of $\psi$ can be calculated by simple algebra:
$$
\begin{array}{ll}
{\rm(ii)}&\psi^x=xF(x,z,\bar z),\quad
\psi^z=zG(x,z,\bar z)+\bar z^{k-1}H(x,z,\bar z)\\
{\rm(iii)}&\psi^x=xF(x,z,\bar z)+z^kJ(x,z,\bar z)+\bar z^k\bar J(x,z,\bar z),
\quad\psi^z=zG(x,z,\bar z)
\end{array}
$$
where $F$ is real, generically $F(0,0,0)\ne0$, $G(0,0,0)\ne0$ and
$J(0,0,0)\ne0$. Thus in cases (i) and (ii),
for small $x$ and $z$ the asymptotically largest terms of $\psi$ are
\begin{equation}\label{mp1}
\psi^x=ax,\quad\psi^z=bz+c\bar z^s,
\end{equation}
where $a$ is real, $b$ and $c$ are complex, and generically $a\ne0$ and $b\ne0$.
(Here $b$ and $c$ are further restricted by the action of other symmetries
from $\Sigma$, but these restrictions are insignificant for the study
of stability of the Poincar\'e map.)
In case (iii) the asymptotically largest terms of the map $\psi$ are
\begin{equation}\label{mp2}
\psi^x=ax+bz^k+\bar b\bar z^k,\quad\psi^z=cz,
\end{equation}
where $a$ is real, $b$ and $c$ are complex, $c^k$ is real,
and generically $a\ne0$, $b\ne0$
and $c\ne 0$. We refer to a cycle with the global map (\ref{mp1}) in cases (i)
or (ii) as a 3-2-[1][23] cycle, and to a cycle with the global map (\ref{mp2})
in case (iii) as a 3-2-[1][23]m cycle.

\section{Stability of the 3-3-[2][1][3] homoclinic cycle}\label{app_2}

In this appendix we derive necessary and sufficient conditions for asymptotic
stability and fragmentary asymptotic stability of the 3-3-[2][1][3] homoclinic
cycle in terms of eigenvalues of the linearisation near the equilibrium.
As noted in section \ref{sec6}, such a cycle is of type Z. Conditions
for stability of type Z cycles in terms of eigenvalues and eigenvectors
of their transition matrices are given by theorems \ref{th_2} and \ref{th_3}.
The transition matrix of the cycle is
\begin{equation}\label{tr1}
M=\left(
\begin{array}{ccc}
a_2&1&0\\
a_3&0&1\\
a_1&0&0
\end{array}
\right)
\end{equation}
(see (\ref{tr33})), where $a_i$ are related to eigenvalues of the linearisation
by (\ref{coeB}), $a_1>0$ ($a_2$ and $a_3$ can have arbitrary signs).

Let $\lambda_1$, $\lambda_2$ and $\lambda_3$ be the eigenvalues of (\ref{tr1});
$\lambda_1$ denotes the largest eigenvalue if all eigenvalues are real, or
the real eigenvalue if the matrix has complex ones. Let $\bf w$ denote
the eigenvector associated with $\lambda_1$. By theorem \ref{th_2},
the necessary and sufficient conditions for asymptotic stability are
\begin{equation}
a_2>0,\ a_3>0,\ \max_j|\lambda_j|>1.
\end{equation}
By theorem \ref{th_3}, the cycle is fragmentarily asymptotically stable
if and only if
\begin{equation}\label{cond41}
\lambda_1>1;
\end{equation}
\begin{equation}\label{cond42}
\lambda_1>\max(|\lambda_2|,|\lambda_3|);
\end{equation}
\begin{equation}\label{cond43}
w_iw_j>0\hbox{ for any }1\le i,j\le3.
\end{equation}

By applying the following lemmas, one can avoid calculating the eigenvalues
$\lambda_i$ by Cardano's formulae for the roots of a cubic polynomial.
\begin{lemma}\label{le1}
Let all $a_i>0$ in matrix (\ref{tr1}). Then $\max_j|\lambda_j|>1$ if and only if
\begin{equation}\label{sum1}
a_1+a_2+a_3>1.
\end{equation}
\end{lemma}

\proof
Eigenvalues of matrix (\ref{tr1}) are roots of its characteristic polynomial
\begin{equation}\label{cha1}
p_M(\lambda)=-\lambda^3+a_2\lambda^2+a_3\lambda+a_1.
\end{equation}

Suppose the inequality (\ref{sum1}) is satisfied. This implies $p_M(1)>0$. Since
$p_M(\infty)<0$, the polynomial $p_M$ has a root larger than one, and thus
$\max_j|\lambda_j|>1$.

We prove now the converse. Denote by $\lambda_{\max}$ the maximal in absolute
value root of $p_M(\lambda)$. Since all $a_i>0$ and $|\lambda_{\max}|>1$,
$$a_1+a_2+a_3>a_2+{a_3\over|\lambda_{\max}|}+{a_1\over|\lambda_{\max}|^2}
\ge\left|a_2+{a_3\over\lambda_{\max}}+{a_1\over\lambda_{\max}^2}\right|=
|\lambda_{\max}|>1.$$
\qed

Components of the eigenvector
$\bf w$ associated with the eigenvalue $\lambda_1$ satisfy the equations
\begin{eqnarray}
a_2w_1+w_2&=&\lambda_1w_1,\label{sys1}\\
a_3w_1+w_3&=&\lambda_1w_2,\label{sys2}\\
a_1w_1&=&\lambda_1w_3.\label{sys3}
\end{eqnarray}
By the Vi\`ete formulae for the roots of the characteristic polynomial $p_M$,
\begin{eqnarray}
\lambda_1+\lambda_2+\lambda_3&=&a_2,\label{vi1}\\
-\lambda_1\lambda_2-\lambda_2\lambda_3-\lambda_1\lambda_3&=&a_3,\label{vi2}\\
\lambda_1\lambda_2\lambda_3&=&a_1.\label{vi3}
\end{eqnarray}

\begin{lemma}\label{lem2}
Eigenvalues and eigenvectors of matrix (\ref{tr1}) satisfy
conditions (\ref{cond41})-(\ref{cond43}) if and only if
the following four inequalities hold true:
\begin{equation}\label{condl2}
\begin{array}{l}
a_1>0,\\
a_1+a_2+a_3>1,\\
a_2a_3+a_1>0,\\
a_1a_2^3+a_3^3>0.
\end{array}
\end{equation}
\end{lemma}

\proof
The cubic polynomial $p_M$ has either three real roots or one real root
and two complex conjugate ones. We consider the two cases separately.

\medskip
Suppose all eigenvalues $\lambda_i$ are real.

Assume that (\ref{cond41})-(\ref{cond43}) hold true. By virtue of
(\ref{sys3}) and (\ref{cond41}), and since $w_1$ and $w_3$ have same signs
(\ref{cond43}), we have $a_1>0$; hence by (\ref{vi3}) $\lambda_2$ and
$\lambda_3$ have same signs. Equations (\ref{vi1}) and (\ref{sys1}) yield
\begin{equation}\label{w1w2}
-w_1(\lambda_2+\lambda_3)=w_2,
\end{equation}
whereby $\lambda_2+\lambda_3<0$ (see (\ref{cond43})). Therefore, $\lambda_2<0$
and $\lambda_3<0$. Thus, (\ref{cond42}) and the identity
\begin{equation}\label{idlem0}
(\lambda_1^2-\lambda_2\lambda_3)
(\lambda_2^2-\lambda_3\lambda_1)(\lambda_3^2-\lambda_1\lambda_2)=a_1a_2^3+a_3^3
\end{equation}
(which follows from the Vi\`ete formulae) yield
$$a_1a_2^3+a_3^3>0.$$
The characteristic polynomial $p_M$ (\ref{cha1}) has only
one positive root that is larger than one. Since $p_M(\infty)<0$, this implies
$p_M(1)=-1+a_1+a_2+a_3>0$. The identity
\begin{equation}\label{idlem}
(\lambda_1+\lambda_2)(\lambda_2+\lambda_3)(\lambda_3+\lambda_1)=-a_2a_3-a_1,
\end{equation}
and (\ref{cond42}) yield $a_2a_3+a_1>0$. Thus, all inequalities in
(\ref{condl2}) are proven.

We prove now the converse assuming that (\ref{condl2}) is satisfied. Since
$a_1>0$, by virtue of (\ref{vi3}) $\lambda_1>0$ for our ordering of $\lambda_i$,
and $\lambda_2\lambda_3>0$. The inequality $a_2a_3+a_1>0$ and (\ref{idlem}) imply
that $\lambda_2<0$ and $\lambda_3<0$. From (\ref{sys3}) we deduce that $w_1$ and
$w_3$ have same signs; by virtue of (\ref{w1w2}) $w_1$ and $w_2$ also have same
signs, which proves (\ref{cond43}). Since $p_M(1)>0$ and $p_M(\infty)<0$,
(\ref{cond41}) holds true. Due to (\ref{idlem0}) and (\ref{idlem}),
the inequalities $a_2a_3+a_1>0$ and $a_1a_2^3+a_3^3>0$ imply that the condition
(\ref{cond42}) holds true.

\bigskip
Suppose now the polynomial $p_M$ has one real root $\lambda_1$ and two
complex conjugate roots $\lambda_{2,3}=\alpha\pm\ri\beta$. Identities
(\ref{idlem}) and (\ref{idlem0}) yield, respectively,
\begin{equation}\label{iden1}
a_2a_3+a_1=-2\alpha((\lambda_1+\alpha)^2+\beta^2)
\end{equation}
and
\begin{equation}\label{iden2}
a_1a_2^3+a_3^3=\Theta(\lambda_1^2-\lambda_2\lambda_3)(\alpha^2+\beta^2)^{-1}
\end{equation}
where
\begin{equation}\label{iden22}
\Theta\equiv(\lambda_1(\alpha^2+\beta^2)-\alpha^3+3\alpha\beta^2)^2
+(3\alpha^2\beta-\beta^3)^2>0,
\end{equation}
unless $a_1+a_2+a_3=0$ and $3\alpha^2=\beta^2$.

By (\ref{iden2}) and (\ref{iden22}), condition (\ref{cond42}) is equivalent
to the inequality $a_1a_2^3+a_3^3>0$.
Since $\lambda_2\lambda_3=|\lambda_2|^2$, (\ref{sys3}) and (\ref{vi3})
imply that $w_1$ and $w_3$ have same signs. By (\ref{iden1}) the inequalities
$a_2a_3+a_1>0$ and $\alpha<0$ are equivalent. By virtue of (\ref{vi1}),
(\ref{sys1}) reduces to
$$-2\alpha w_1=w_2,$$
and hence $a_2a_3+a_1>0$ is equivalent to (\ref{cond43}).

For the characteristic polynomial $p_M$ (\ref{cha1}) with only one real
eigenvalue the conditions $(\ref{cond41})$ and $a_1+a_2+a_3>1$ are equivalent.
Finally, (\ref{vi3}) imply that $a_1$ and $\lambda_1$ have same signs.
\qed

Substituting expressions (\ref{coeB}) into the inequalities in the statements
of lemmas \ref{le1} and \ref{lem2}, we establish conditions for stability
of the cycle presented in table 2.

\section{Stability of the 3-2-[1][23]m homoclinic cycle}\label{app_3}

In this appendix we derive necessary and sufficient conditions for asymptotic
stability of the 3-2-[1][23]m homoclinic cycle. The Poincar\'e map near the cycle is
\begin{equation}\label{mapa1}
g(w,z)=(Aw^{a_1}+|w|^{a_2k}{\rm Re}(bz^k),cz|w|^{a_2})
\end{equation}
where $z=z_1+\ri z_2$, $A$ is real, $B$ and $c$ are complex and $c^k$ is real
(see table \ref{tab3} and the expression for the global map (\ref{mp2}) in
appendix \ref{app_1}). Recall that $a_1=-c/e$, and therefore by definition
of expanding and contracting eigenvalues (see subsection \ref{hetc}) $a_1$ is
always positive. Upon the coordinate transformation $bz^k=u+\ri v$, (\ref{mapa1})
becomes
\begin{equation}\label{mapa2}
g(w,u,v)=(Aw^{a_1}+|w|^{a_2k}u,Cu|w|^{a_2k},Cv|w|^{a_2k})
\end{equation}
where $C=c^k$ is real.

For $a_1>a_2k$ and $a_1<1$, we partition $\R^3$ into two regions (if $a_1<a_2k$
or $a_1>1$, the partitioning is not needed, see the proof of lemma \ref{la6}):
\begin{equation}\label{deco}
\renewcommand{\arraystretch}{1.8}
\begin{array}{lll}
\Omega_I&=&\{(w,u,v):|Aw^{a_1}+|w|^{a_2k}u|<|w^{\alpha}|\}\\
\Omega_{II}&=&\R^3\setminus\Omega_I,
\end{array}
\end{equation}
where $\alpha$ satisfies the inequality $a_1<\alpha<1$.

Denote by $B_{\epsilon}$ the $\epsilon$-neighbourhood of the point
$(w,u,v)=\bf 0$, and $\Omega_J(\epsilon)=\Omega_J\cap B_{\epsilon}$ for
$J=I,II$. Also,
$$f_1(x)\approx f_2(x)\hbox{ denotes that }f_1(x)-f_2(x)=\ro(f_1(x))
\hbox{ when }x\to0$$
and
$$f_1(x)\sim f_2(x)\hbox{ denotes that }f_1(x)\approx Ff_2(x)
\hbox{ for a constant }F\ne 0.$$
Finally, denote
\begin{equation}\label{defQ}
Q(f_1,f_2)=\{(w,u,v):\ f_1(w)<u<f_2(w)\}\hbox{ and }H(f)=\{(w,u,v):\ u=f(w)\}.
\end{equation}
Let $f_0(w)$ be the solution to $Aw^{a_1}+|w|^{a_2k}f_0(w)=0$, $l_0=H(f_0)$
and $l_{j+1}=g^{-1}l_j$ be the preimage of $l_j$ under $g$.

\begin{lemma}\label{lem_aux1}
Suppose $Q(d_1,d_2)=gQ(f_1,f_2)$, where
$$f_j=f_0+h_j,\ h_j\approx Dw^{a_1/(a_1-a_2k)-a_2k},\hbox{ for }
j=1,2,\ h_1-h_2\sim w^{\beta},\ \beta>a_1/(a_1-a_2k)-a_2k.$$
Then
$$d_1-d_2\sim w^{\beta+a_1-a_1/(a_1-a_2k)}.$$
\end{lemma}

\proof
By the condition of the lemma,
\begin{equation}\label{la1}
g(w,f_j(w),u)\approx(w^{a_2k}h_j,E_1w^{a_1},\tilde u)\approx
(E_2w^{a_1/(a_1-a_2k)},E_1w^{a_1},\tilde u),
\end{equation}
\begin{equation}\label{la2}
\begin{array}{ll}
g(w,f_2(w),u)\approx&(w^{a_2k}h_1+w^{a_2k}(h_2-h_1),E_1w^{a_1},\tilde u)\approx\\
&\left(((w^{a_2k}h_1)^{(a_1+a_2k)/a_1}+E_3w^{\beta+a_2k+a_2k/(a_1-a_2k)})^{a_1/(a_1-a_2k)},
E_1w^{a_1},\tilde u \right).
\end{array}\end{equation}
(Note that $(w^{a_2k}h_1)^{(a_1+a_2k)/a_1}\sim w$.)
For small $|w_1-w_2|$,
$$d_1(w_1)-d_2(w_2)=d_1(w_1)-d_2(w_1)+d_2(w_1)-d_2(w_2)\approx
d_1(w_1)-d_2(w_1)+d_2'(w_1)(w_1-w_2).$$
Set $w_1=(w^{a_2k}h_1)^{(a_1+a_2k)/a_1}$ and $w_2=w_1+E_3w^{\beta+a_2k+a_2k/(a_1-a_2k)}$.
By (\ref{la1}), \hbox{$d_2'(w_1)\sim w^{a_1-a_2k-1}$} and $d_1(w_1)=d_2(w_2)$.
Therefore, $d_1(w)-d_2(w)\sim w^{\delta}$, where \hbox{$\delta=\beta+a_1-a_1/(a_1-a_2k)$}.
\qed

\begin{lemma}\label{la6}
Consider the map
\begin{equation}\label{mapl11}
g(w,u,v)=(Aw^{a_1}+|w|^{a_2k}u,Cu|w|^{a_2k},Cv|w|^{a_2k}),
\end{equation}
where $a_1>0$.

\begin{itemize}
\item[(i)]
If
\begin{equation}\label{ineq}
a_1>1\hbox{ and }a_2>0,
\end{equation}
then the fixed point $(w,u,v)=\bf 0$ of the map $g$ is asymptotically stable.

\item[(ii)]
If
$$a_1<1\hbox{ or }a_2<0,$$
then the fixed point $(w,u,v)=\bf 0$ of the map $g$ is completely unstable.
\end{itemize}
\end{lemma}

\proof
(i) Suppose $a_1>1$ and $a_2>0$. Then the iterates
$(w_{j+1},u_{j+1},v_{j+1})=g(w_j,u_j,v_j)$ satisfy the inequalities
$$|w_{j+1}|<\max(2|Aw_j^{a_1}|,2|w_j|^{a_2k}|u_j|),\ C|u_{j+1}|<|w_j|^{a_2k}|u_j|,\
|v_{j+1}|<C|w_j|^{a_2k}|v_j|.$$
Therefore, if $|(w_j,u_j,v_j)|<\epsilon$, where $A\epsilon^{a_1}<1/4$ and
$\max(|C|,1)\epsilon^{a_2k}<1/4$, then\break$|w_{j+1},u_{j+1},v_{j+1})|<|(w_j,u_j,v_j)|$
and $\lim_{j\to\infty}|(w_j,u_j,v_j)|=0$.

\medskip
(ii) Suppose $a_2<0$. Since $u_{j+1}=Cu_j|w|^{a_2k}$ for $C\epsilon^{a_2k}>1$,
the iterates $(w_j,u_j,v_j)$ escape from $B_{\epsilon}$, unless
$(w_0,u_0,v_0)\in l_k$ for some $k\ge 0$. The measure of the union of the sets $l_k$
is zero. This implies the statement of the lemma.

\medskip
Suppose now $a_2>0$ and $a_1<1$. Consider the sets
$${\cal H}_j=\cup_{j\le l\le\infty} g^{-l}\Omega_I,\
{\cal H}_j(\epsilon)={\cal H}_j\cap B_{\epsilon}.$$
Let $\mu$ denote the Lebesgue measure in $\R^3$.
Since $\Omega_I=Q(f_1,f_2)$ where $f_1-f_2\sim w^{\alpha-a_2k}$, we have
$\mu(\Omega_I(\epsilon))\sim\epsilon^{2+(\alpha-a_1+1)/(a_1-a_2k)}$.
By lemma \ref{lem_aux1},
$g^{-l}\Omega_I=Q(d_1,d_2)$ where $d_1-d_2\sim w^{\alpha-a_2k+ls}$ and
$s=a_1/(a_1-a_2k)-a_1>0$. Therefore,
$\mu(g^{-l}\Omega_I(\epsilon))\sim\epsilon^{2+(\alpha-a_1+1+sl)/(a_1-a_2k)}$.
Thus, $\lim_{j\to\infty}\mu({\cal H}_j(\epsilon))=0$.

If $(w_0,u_0,v_0)\notin{\cal H}_j$ and $(w_0,u_0,v_0)\notin\cup_{0\le k\le\infty}l_k$,
then $w_j\ne0$ and $|w_{l+1}|>|w_l|^{\alpha}$ for $l\ge j$. Since $\alpha<1$,
this indicates that almost all $(w_0,u_0,v_0)$ (except for a set of zero
measure) escape from $B_{\epsilon}$ for a sufficiently small $\epsilon>0$.
\qed
\end{document}